\documentclass[useAMS,usenatbib]{mn2e}
\usepackage{mathrsfs} 
\usepackage[dvipdfmx]{graphicx}
\usepackage{mediabb}
\usepackage{color}

\def\Msun{M_{\odot}}   
\def\r{\bibitem[]{}}  
\def\kms{km s$^{-1}$}     
\def\Htwo{H$_2$} \def\H2{H$_2$}

\def\rhoh2{\rho_{\rm H_2}} \def\rhohi{\rho_{\rm HI}} 
\def\rhotot{\rho_{\rm tot}} 
\def\vsfr{\rho_{\rm SFR}} \def\vsfr{\rho_{\rm SFR}}  

\def\vsfr{\rho_{\rm SFR}}  
\def\rhogas{\rho_{\rm gas}} 
\def\vsfrunitkpc{\Msun~{\rm kpc^{-3}y^{-1}}} 
\def\vsfrunit{\Msun~{\rm pc^{-3}Gy^{-1}}} 
\def\vsfrunitGy{\Msun~{\rm pc^{-3}Gy^{-1}}} 
\def\alphah2{\alpha_{\rm H_2}} 
 
  \def\vmu{{\rm H\ pc^{-3}}} 
\def\vgas{\rho_{\rm gas}} 
  
\def\log{{\rm log}} \def\hcc{{\rm H~cm^{-3}}} 
\def\Hcc{$\hcc$}  
\def\vr{v_{\rm r}} \def\A{A_{\rm vol}}
\def\sigtot{\Sigma_{\rm tot}} \def\sigh2{\Sigma_{\rm H_2}} \def\sighi{\Sigma_{\rm HI}}  \def\ssfr{\Sigma_{\rm SFR}} 
\def\ssfr{\Sigma_{\rm SFR}} \def\sgas{\Sigma_{\rm gas}} 
\def\sfu{\Msun~{\rm y^{-1}~kpc^{-2}}}  
\def\smu{\Msun~{\rm pc^{-2}}}

\title[Variable Star Formation Law in the Galaxy]{Radial Variations of the Volume- and Surface-Star Formation Laws in the Galaxy}  
\author[Yoshiaki SOFUE]{Yoshiaki \textsc{Sofue}$^{1}$\thanks{E-mail: sofue@ioa.s.u-tokyo.ac.jp}\\
$^1$Insitute of Astronomy, The University of Tokyo, Mitaka, Tokyo 181-0015, Japan }   

\begin{document} 
\date{ }

\maketitle  

\begin{abstract}  
Variation of the volume- and surface-Schmidt laws (star-formation or SF law) with the galacto-centric distance $R$ was investigated using 3D distributions of HII regions, HI, and molecular (\Htwo) gases in the Milky Way. Both the power-law index and SF coefficient were found to be variable with $R$. The index is flatter in the inner disc than in the outer Galaxy, and the coefficient is larger in the inner disc, decreasing steeply outward. There is also a mutual anti-correlation between the index and SF coefficient, and the SF law can be expressed by a single-parameter function of the SF coefficient. The variable SF law is discussed in relation to the self-regulation star formation.  
\end{abstract}  
 

\section{INTRODUCTION} 
 
The Schmidt-Kennicutt law (Schmidt 1959; Kennicutt 1998a, b), hereafter star formation or SF law, in galaxies has been extensively investigated by measuring surface densities of star formation rate (SFR) and surface or column gas densities projected on the galactic planes (Schmidt 1959; Kennicutt 1998a, b; Komugi et al. 2006; Bigiel et al. 2008; Leroy et al. 2008, 2013; Lada et al. 2012; Kennicutt and Evans 2012; and the large body of literature therein). It is established that the SF law represents the universal scaling relation in galactic-scale star formation. Detailed analyses of dependence of the law on the ISM condition, particularly on the molecular gas have been recently obtained in several galaxies, and the molecular-gas regulated star formation scenario is proposed and modeled (Boissier et al. 2003; Booth et al. 2007; Leroy et al. 2008; Krumholz et al. 2009, 2013).  

The SF law in the Milky Way has been investigated using individual star forming regions associated with molecular and HI clouds and spiral arms (Fuchs et al. 2006; Luna et al. 2006; Heiderman et al. 2010; Willis et al. 2015; Heyer et al. 2016; and the papers cited therein).
    Recently, we analyzed the Galactic distributions of HII regions, \Htwo, and HI gases to obtain a global SF law in the entire Milky Way with spatial resolution of $\sim 1$ kpc in the Galactic plane (Sofue and Nakanishi 2016b; Paper I). 
    
We showed that HII regions are distributed in a star-forming disc with nearly constant full thickness of 92 pc in spatial coincidence with the molecular gas disc. With this thickness, the vertically averaged volume star formation rate (SFR), $\vsfr$, was shown to be related to the surface SFR, $\ssfr$, by 
$\vsfr / [\vsfrunitkpc] = \vsfr / [\vsfrunitGy] = 9.26 \times \ssfr / [\sfu]$.
This relation will be used throughout this paper.
  
In this paper, we examine the variations of both the volume and surface SF laws in the Galaxy. The volume SF law in the whole Galactic disc will be analyzed in detail for the first time, which can be obtained uniquely by analyzing the 3D data of HI and \Htwo\ gas distributions. Both the volume- and surface-SF laws will be shown to be not universal in the Galaxy, but highly variable with the galacto-centric distance and dependent on the SFR itself. 

In the analyses we use the HII region catalogue in the Milky Way (Hou and Han 2014) and the 3D maps of HI and \Htwo\ gases (Nakanishi and Sofue 2005, 2006, 2016; Sofue and Nakanishi 2016a), which were obtained by analyzing the Leiden-Argentine-Bonn All Sky HI survey (Kalberla et al. 2005) and Galactic disc CO line survey (Dame et al. 2001). 

\section{Definition of the Volume- and Surface SFR and the Data}
 
Following the definition adopted in Paper I, we express the volume SFR by power-law function of gas density, and represent it in logarithmic form as 
\begin{equation} 
\log \left(\vsfr \over \vsfrunitGy \right)=\alpha\ {\rm log}\left(\rho_{\rm gas} \over {\rm H\ cm^{-3}}\right)+A_{\rm vol},
\label{eqschrho}
\end{equation}
where $\rho_{\rm SFR}$ is the vertically averaged volume SFR in the SF disc of full thickness 92 pc, and $A_{\rm vol}=A+0.967$ with $A$ being the coefficient between $\Sigma_{\rm SFR}$ and $\rho$ defined in Paper I. The coefficient $\A$ represents the $\vsfr$ at $\rhogas=1\vmu$.  
Similarly, the surface SFR law is expressed by
\begin{equation}
\log \left(\ssfr \over \sfu \right) =\beta\ {\rm log} \left(\Sigma_{\rm gas}\over {\rm \Msun~pc^{-2}}\right)+B,
\label{eqschsigma}
\end{equation}
where the coefficient $B$ represents $\ssfr$ at $\sgas=1\smu$. 

The SFR was calculated using the emission rates of UV photons of HII regions inferred from their excitation parameter observed by radio observations (Schraml and Mezger 1969; Hou and Han 2014). The incompleteness of mapping of HII regions was corrected for the detection probability, which is a function of the distance and intrinsic luminosity of each HII region. The distance errors are mostly kinematical distance errors due to intrinsic velocity dispersion of HII regions, while recombination line velocity measurements had better resolution. They are on the same order as those for the HI and \Htwo\ gases, $\sim 0.5$ to 1 kpc, as described below. The number of effective independent data points in each plot of SF law is the number of $\sim 1\times 1$ kpc grids in the Galactic plane, which is taken into account for calculating the standard errors of the plotted quantities.
  
The map of SFR was obtained by smoothing the UV emission rates by a Gaussian beam of FWHM 1 kpc ($\pm 0.5$ kpc) in the Galactic plane. 
The vertical resolution depend on the telescope resolution and positioning, and was of the order of several arc minutes, or $\sim 7$ pc for $\delta \theta \sim 5'$ at $\sim 5$ kpc distance, and was better than that for the gases as below.  However, since the number of HII regions was not sufficient in order to divide the disc into a meaningful number of layers, we do not discuss the vertical variation. 

 The distribution maps of HI and \Htwo\ gases in the Galaxy have spatial resolution depending on their errors in kinematical distances and therefore on the distance from the Sun and direction. Except for the velocity-degenerated regions in the Galactic Center and anti-Center directions as well as along the solar circle, the distance error is about $\sim 0.5$ to 1 kpc for an effective radial velocity uncertainty corresponding to the interstellar velocity dispersion of $\delta \vr \sim 5$ \kms (see Eq. 23 of Sofue (2011)), although the observational resolution of the LAB HI survey was $1.3$ \kms. 

\section{Radial Variation of Star Formation Law}  

We examined how the SF relation varies with the galacto-centric distance $R$. We obtained the same plots both for the volume and surface SF laws, and found the same or similar behaviors. 

\subsection{Non-universal SF Law varying with Radius}

\begin{figure*} 
\begin{center}  
\includegraphics[width=7cm]{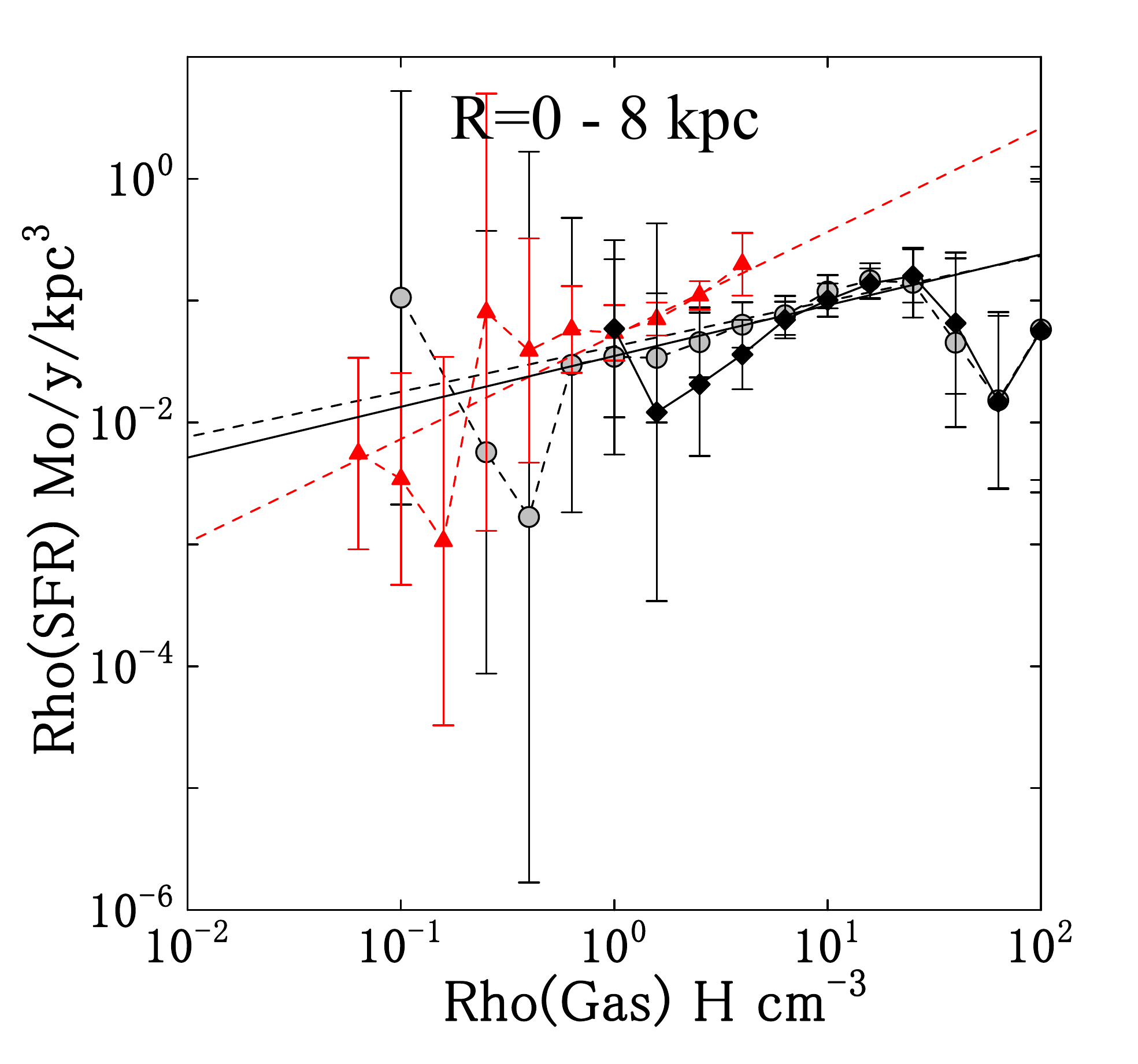}  
\includegraphics[width=7cm]{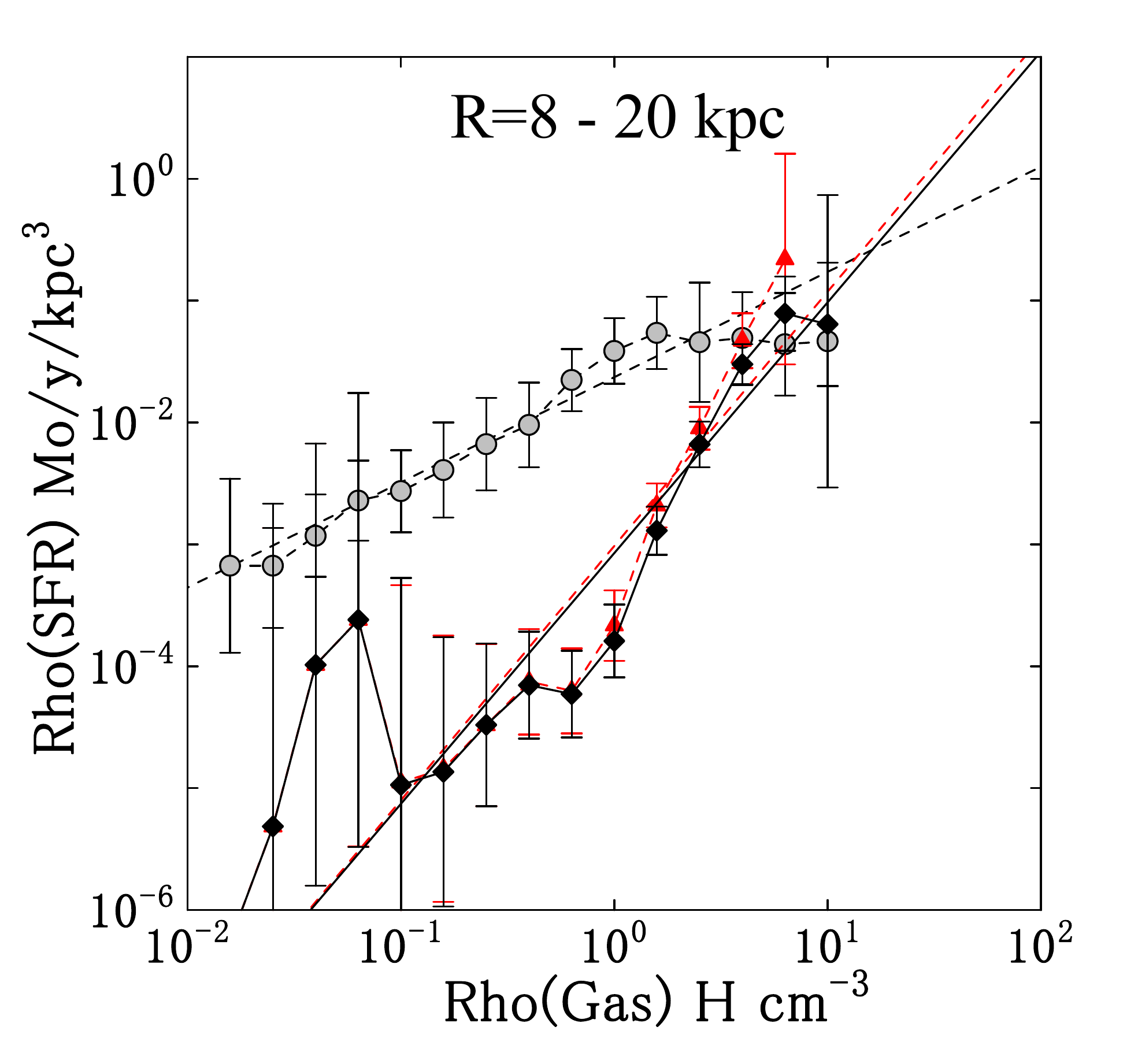}  
\end{center}
\caption{ 
Inner  ($R=0$ to 8 kpc) and outer (8 to 20 kpc) volume SF laws in the Galaxy. Grey big circles are molecular gas, triangles for HI, and diamonds for total (HI + \Htwo) gas. The straight lines are the least-squares fit to the plots. The unit of SFR is equivalent to $\vsfrunitGy$.}
\label{rho8kpc}   

\begin{center}     
\includegraphics[width=7cm]{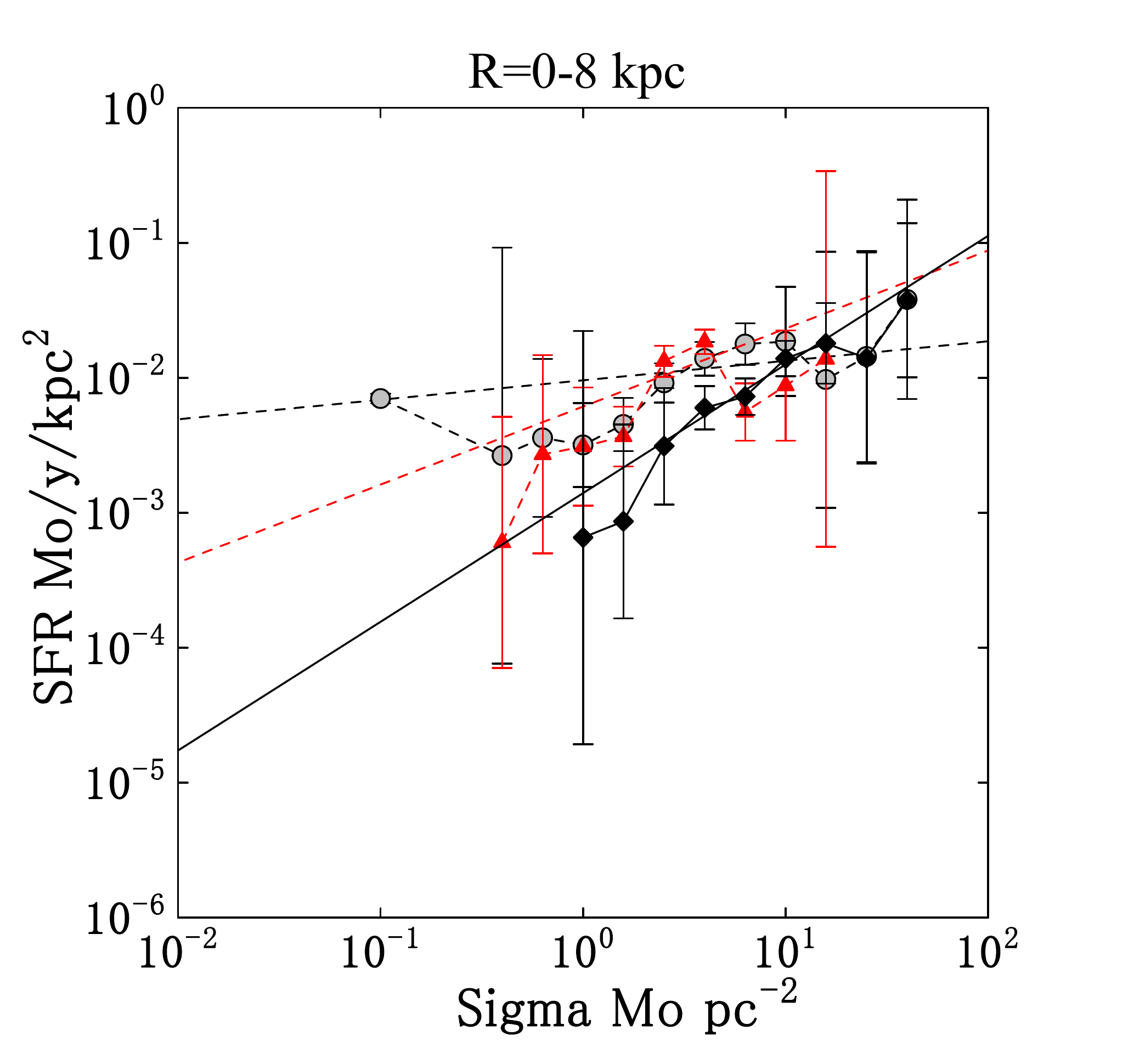}  
\includegraphics[width=7cm]{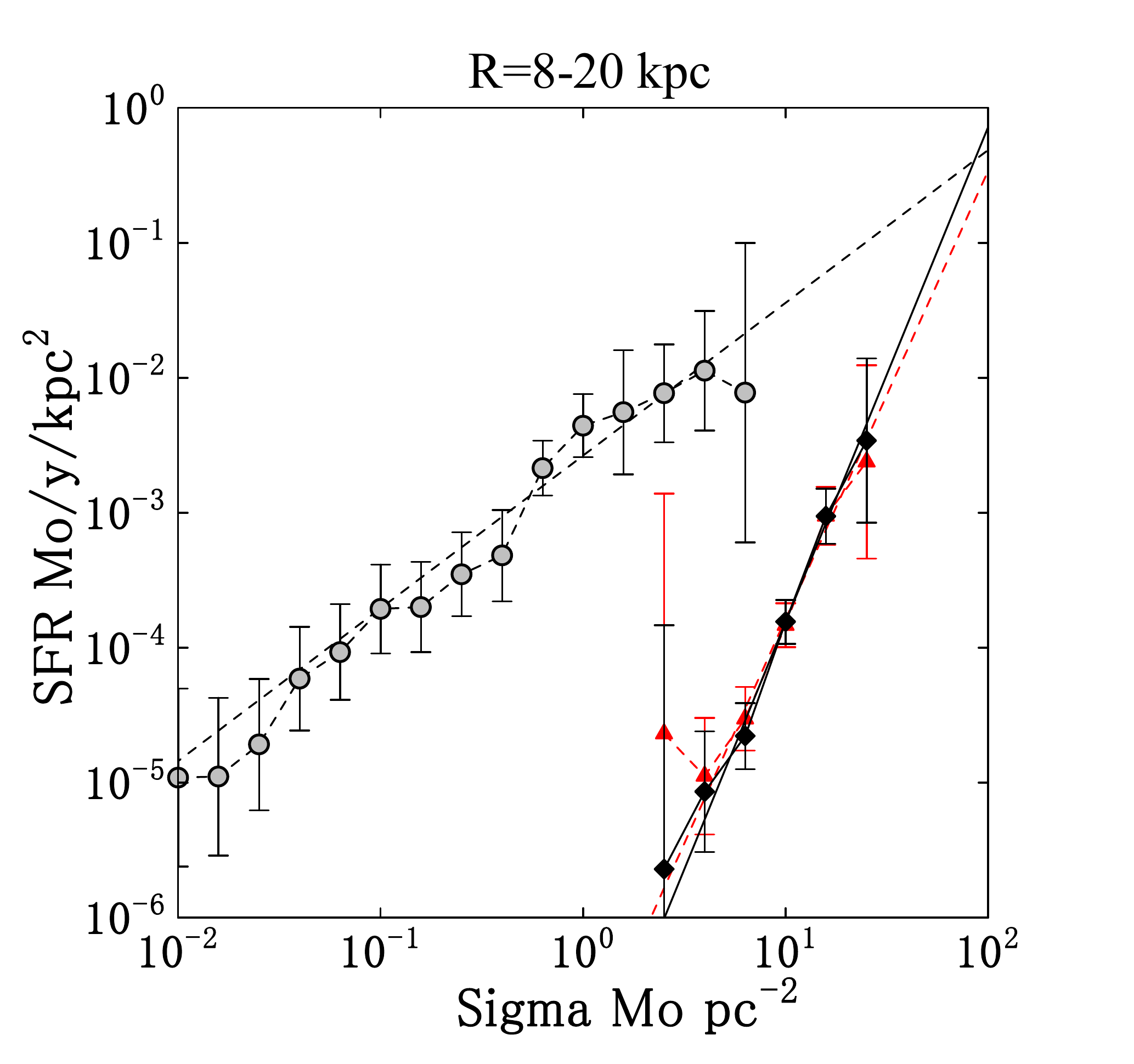}  
\end{center}
\caption{Same as figure \ref{rho8kpc}, but for surface (integrated column) densities. } 
\label{sig8kpc}  
\end{figure*}

First, we divided the Galaxy into two areas of the inner disc at $R\le 8$ kpc, and outer disc at $R> 8$ kpc, and derived the SF law in each area. Figures \ref{rho8kpc} and \ref{sig8kpc} show the results, where the straight lines are the least-squares fitting to the plots in the log-log plane. Table \ref{tab1} and \ref{tab2} list the best-fit values for the indices and coefficients. In the first row of the tables, we list the parameters for the whole Galaxy obtained in Paper I for comparison.

There is remarkable difference in the SF laws in the inner and outer discs. The inner SF law shows significantly smaller power-law index compared to that in the outer disc. Also, the average value of SFR is greater in the inner disc, trivially reflecting the fact that the SF activity is higher in the inner disc because of the higher gas density. 

In order to examine finer variations with $R$, we further divide the Galaxy into a larger number of rings of 2 kpc width. The volume- and surface SF laws made in individual rings are shown in figures \ref{rho2kpc} and \ref{sig2kpc}, respectively, in the Appendix. We fitted each plot by the least-squares fitting, and show the result by inserted straight lines. The best-fit values are listed in tables \ref{tab1} and \ref{tab2}.
 
There are shorter-scale variations of the SF law with the radius. Besides the generally flat index in the inner Galaxy, the flattest, or an even inverse index of the volume SF law was found at $R=2-4$ kpc. On the other hand, the surface SF law exhibits the flattest index at $R=4-6$ kpc, and the innermost and Galactic Center regions show again steep index.

The inverse volume SF law at $R=2-4$ kpc indicates that the SFR decreases with increasing gas density. This apparently strange behavior could be understood, if some regulation mechanism is working in such a way that the SF is strongly suppressed, when the SFR exceeds a certain threshold value. In the present case, this threshold exists at $\vsfr \sim 0.1\ \vsfrunit$. It must be also emphasized that the region coincides in position with the 4-kpc molecular ring with the highest SFR in the Galaxy.

Another remarkable nature of the SF laws at $R<\sim 8$ kpc is the coincidence of both the index and coefficient for \Htwo\ with those for HI and total gases. This may reflect the facts that the inner Galactic disc is dominated by the molecular gas and that the HI gas density is saturated at a threshold density related to the molecular gas density (Sofue and Nakanishi 2016a).

In contrast, the SF law for \Htwo\ gas in the outer Galaxy at $R> \sim 8$ kpc is significantly displaced from those for HI and total gases, showing a much milder index. Also, the outer SF laws are similar to that obtained in the entire Galactic disc in Paper I. This means that the global SF law is more determined by the laws in the outer Galaxy for the larger area, and hence, for the larger weight in the averaging.

Thus, the global SF law, often regarded as the universal scaling relation, reflects more the SF relations in the outer disc. This sounds paradoxical in the sense that the universal SF law may not necessarily be the direct measure of the most active star formation in the galaxy. Namely, the most active SF in the galaxy cannot be investigated by the analysis of the global SF law. 

This paradox reminds us of the SF law in starburst galaxies, which is significantly displaced from the universal SF law for normal galaxies (Komugi et al. 2005, 2006). Such displacement could be due to selectively observed active SF regions in starbursts, having intrinsically larger SF coefficients.

\begin{table*}
\caption{Index $\alpha$ and coefficient $\A$ for volume-SF law at different $R$ ranges.}
\label{tab1} 
\begin{center}
\begin{tabular}{lllllll}  
\hline
\hline
$R$ (kpc) & $\alpha$(Mol) & $\A$(Mol) & $\alpha$(HI) & $\A$(HI) & $\alpha$(Tot) & $\A$(Tot) \\ 
\hline 
0.00 --  20.00
 &  0.70$\pm$   0.07& -1.69$\pm $  0.07
 &  2.29$\pm$   0.03& -2.59$\pm $  0.06
 &  2.01$\pm$   0.02& -3.17$\pm $  0.04
\\ 
\hline 
   0.00 --   8.00
 &  0.37$\pm$   0.17& -1.38$\pm $  0.18
 &  0.85$\pm$   0.27& -1.28$\pm $  0.10
 &  0.42$\pm$   0.20& -1.45$\pm $  0.22
\\
   8.00 --  16.00
 &  0.86$\pm$   0.14& -1.63$\pm $  0.11
 &  2.09$\pm$   0.04& -3.00$\pm $  0.08
 &  2.06$\pm$   0.04& -3.06$\pm $  0.07
\\
\hline 
   0.00 --   2.00
 &  0.61$\pm$   0.28& -2.40$\pm $  0.48
 &  0.85$\pm$   0.41& -1.79$\pm $  0.26
 &  0.43$\pm$   0.13& -2.07$\pm $  0.04
\\
   2.00 --   4.00
 & -0.73$\pm$   0.24& -0.56$\pm $  0.20
 & -0.72$\pm$   0.43& -1.10$\pm $  0.17
 & -0.80$\pm$   0.21& -0.41$\pm $  0.17
\\
   4.00 --   6.00
 &  0.33$\pm$   0.19& -1.00$\pm $  0.20
 &  0.52$\pm$   0.35& -0.82$\pm $  0.11
 &  0.23$\pm$   0.10& -0.90$\pm $  0.09
\\
   6.00 --   8.00
 &  0.41$\pm$   0.28& -1.40$\pm $  0.25
 &  0.78$\pm$   0.53& -1.38$\pm $  0.18
 &  0.55$\pm$   0.30& -1.60$\pm $  0.30
\\
   8.00 --  10.00
 &  0.39$\pm$   0.20& -1.43$\pm $  0.11
 &  2.85$\pm$   0.55& -2.55$\pm $  0.26
 &  1.91$\pm$   0.50& -2.45$\pm $  0.29
\\
  10.00 --  12.00
 &  0.42$\pm$   0.17& -2.31$\pm $  0.13
 &  2.33$\pm$   0.58& -3.41$\pm $  0.24
 &  1.85$\pm$   0.47& -3.35$\pm $  0.24
\\
  12.00 --  14.00
 &  1.82$\pm$   0.68& -3.66$\pm $  0.20
 &  1.83$\pm$   0.68& -3.66$\pm $  0.20
\\
  14.00 --  16.00
 &  1.80$\pm$   0.20& -3.52$\pm $  0.40
 &  1.91$\pm$   0.20& -3.30$\pm $  0.40
\\
  16.00 --  18.00
 &  1.91$\pm$   0.20& -3.30$\pm $  0.40
\\
\hline  
\end{tabular}
\end{center}

\caption{Index $\beta$ and coefficient $B$ for surface-SF law at different $R$ ranges.}
\label{tab2} 
\begin{center}
\begin{tabular}{lllllll}  
\hline
\hline
$R$ (kpc) & $\beta$(Mol) & $B$(Mol) & $\beta$(HI) & $B$(HI) & $\beta$(Tot) & $B$(Tot) \\ 
\hline 
Galaxies$^\dagger$
 &  ---&  ---
 &  --- & ---
 &  1.4 &  $-3.6$
\\
\hline 
MW\\
   0.00 --  20.00
 &  1.10$\pm$   0.07& -2.57$\pm $  0.06
 & -1.13$\pm$   0.21& -2.25$\pm $  0.16
 &  1.12$\pm$   0.37& -4.30$\pm $  0.36
\\
\hline 
   0.00 --   8.00
 &  0.14$\pm$   0.05& -2.02$\pm $  0.05
 &  0.58$\pm$   0.31& -2.21$\pm $  0.17
 &  0.95$\pm$   0.32& -2.85$\pm $  0.28
\\
   8.00 --  16.00
 &  1.13$\pm$   0.11& -2.58$\pm $  0.11
 &  3.13$\pm$   0.60& -6.91$\pm $  0.61
 &  3.45$\pm$   0.59& -7.24$\pm $  0.60
\\ 
\hline 
   0.00 --   2.00
 &  1.43$\pm$   0.54& -3.82$\pm $  0.54
 &  1.68$\pm$   0.90& -3.00$\pm $  0.32
 &  1.61$\pm$   0.57& -4.14$\pm $  0.63
\\
   2.00 --   4.00
 &  0.81$\pm$   0.73& -2.60$\pm $  0.36
 &  1.00$\pm$   0.73& -2.42$\pm $  0.26
 &  0.98$\pm$   0.86& -2.91$\pm $  0.61
\\
   4.00 --   6.00
 &  0.50$\pm$   0.20& -1.94$\pm $  0.12
 &  0.51$\pm$   0.27& -1.89$\pm $  0.13
 &  1.04$\pm$   0.34& -2.55$\pm $  0.29
\\
   6.00 --   8.00
 &  0.09$\pm$   0.07& -2.06$\pm $  0.07
 & -0.32$\pm$   0.47& -1.86$\pm $  0.30
 &  0.39$\pm$   0.59& -2.49$\pm $  0.50
\\
   8.00 --  10.00
 &  0.40$\pm$   0.22& -2.37$\pm $  0.12
 &  0.69$\pm$   0.48& -3.12$\pm $  0.48
 &  2.23$\pm$   0.69& -4.80$\pm $  0.74
\\
  10.00 --  12.00
 &  0.66$\pm$   0.26& -3.14$\pm $  0.24
\\
  12.00 --  14.00
 &  1.00$\pm$   0.33& -3.03$\pm $  0.49
\\
  14.00 --  16.00
 & -1.23$\pm$   0.42& -7.90$\pm $  0.81
\\ 
\hline  
\end{tabular}
\end{center}
\leftline{$\dagger$ Kennicut (1998a)}
\end{table*}

\subsection{Variable SF coefficients and indices }

\begin{figure*} 
\begin{center}       
\includegraphics[width=6cm]{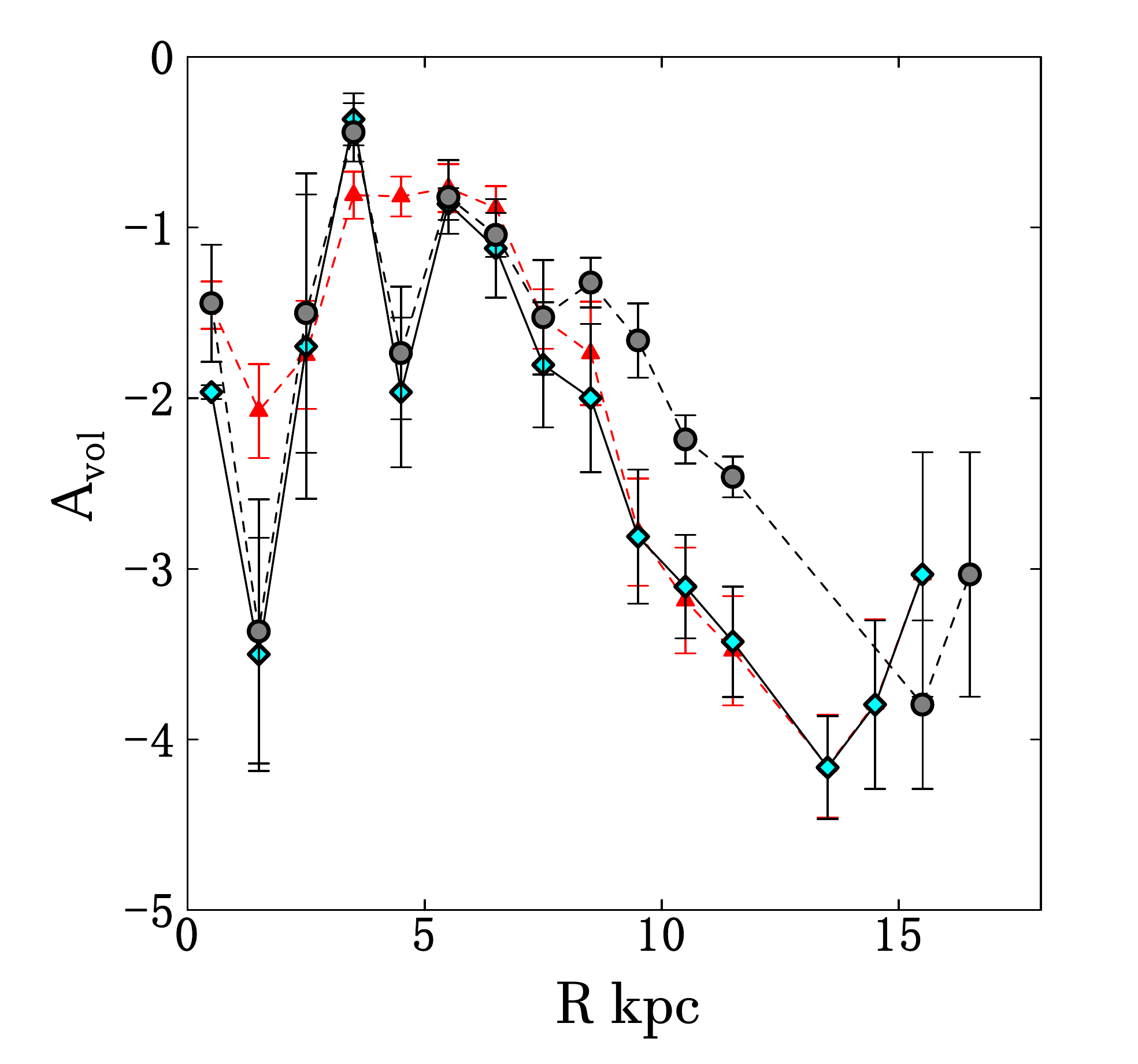}   
\hskip-3mm \includegraphics[width=6cm]{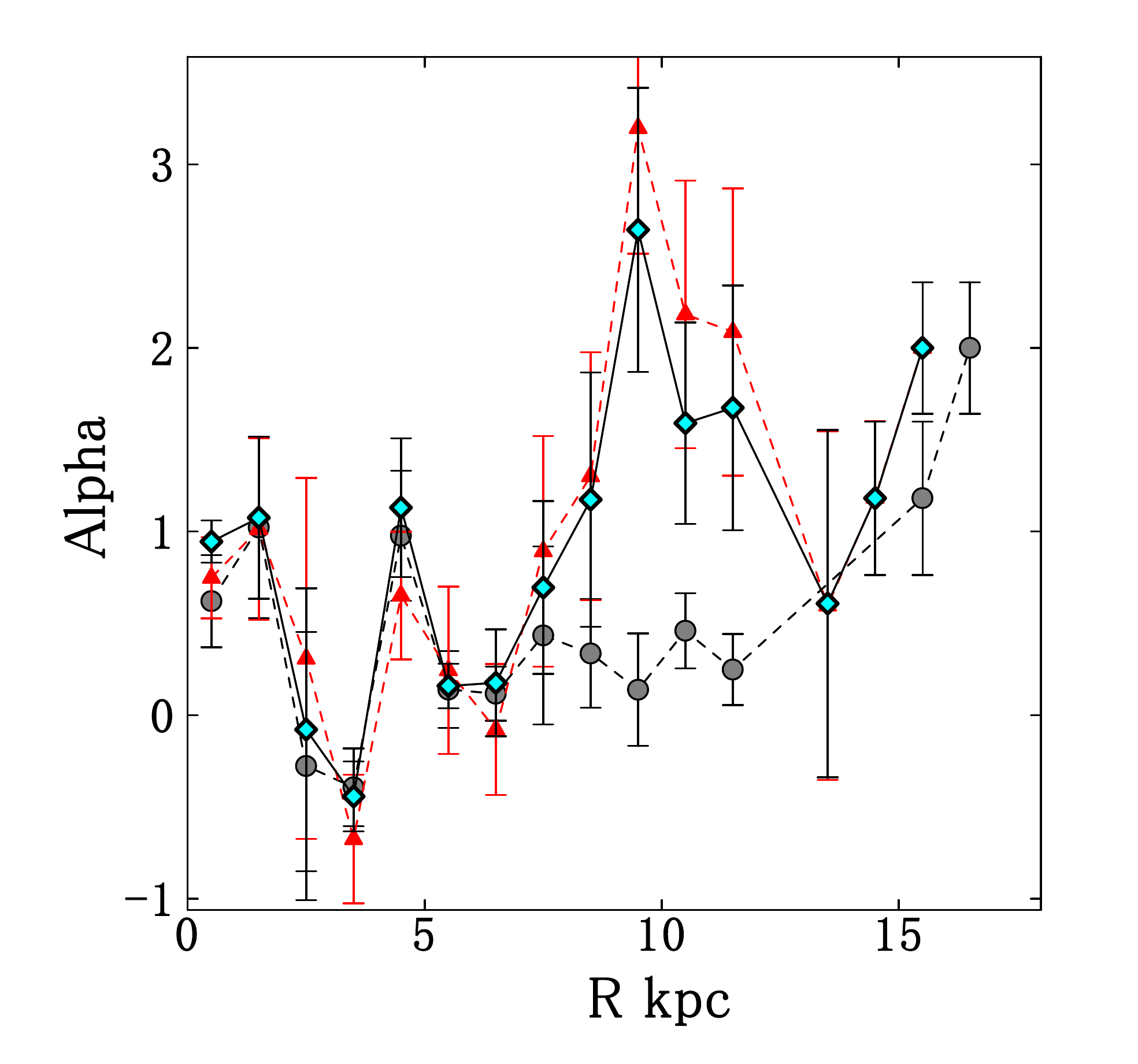} 
\hskip-3mm \includegraphics[width=6cm]{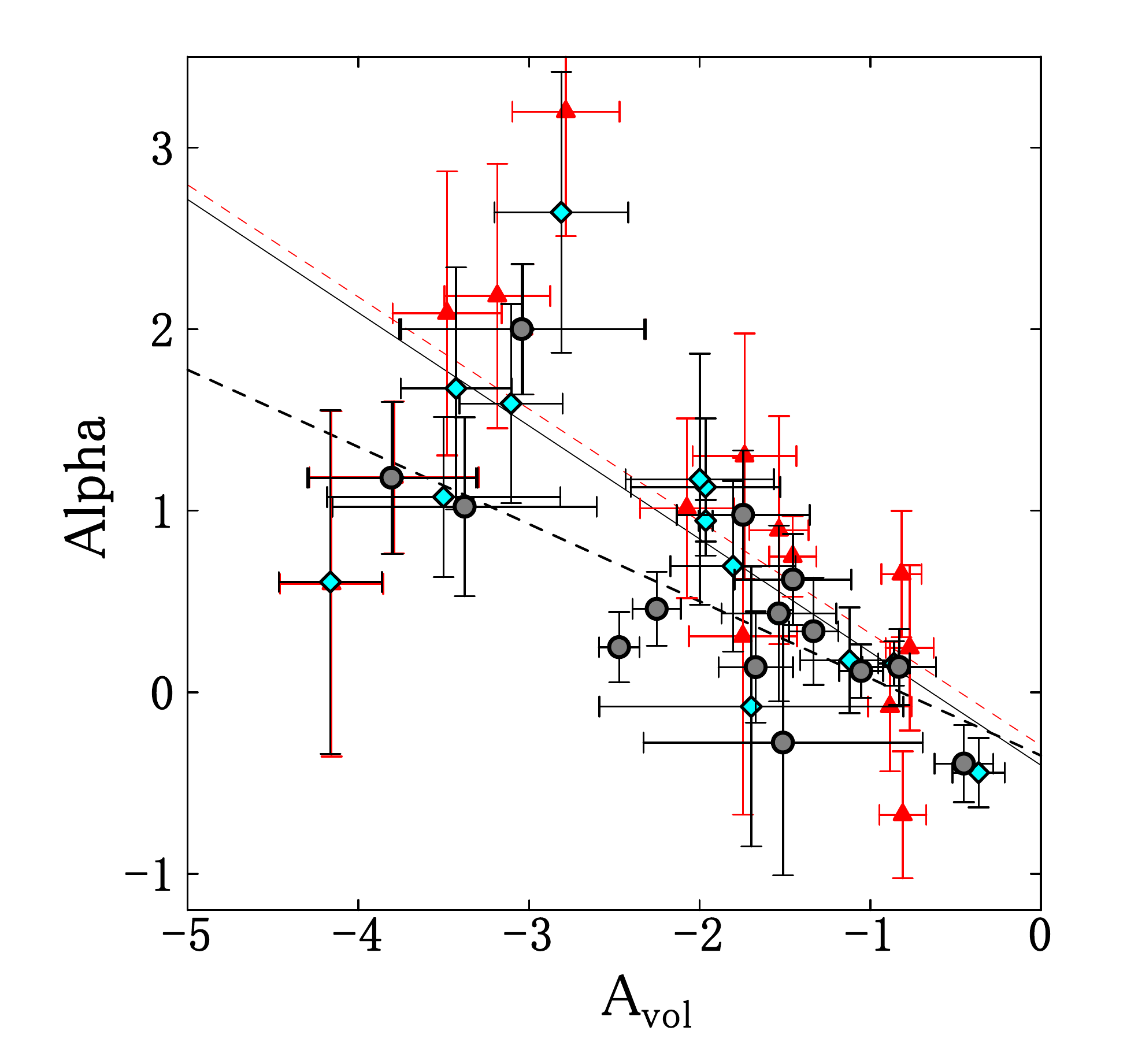}  
\end{center}
\caption{[Left] Radial variation of $\A$. [Middle] Same for $\alpha$.  [Right] Correlation between $\alpha$ and $\A$. Lines show the least-squares fits to individual plots (dash for \Htwo; thin dash for HI; solid for total).  Symbols are same as before.  } 
\label{R-A-alpha} 

\begin{center}   
\includegraphics[width=6cm]{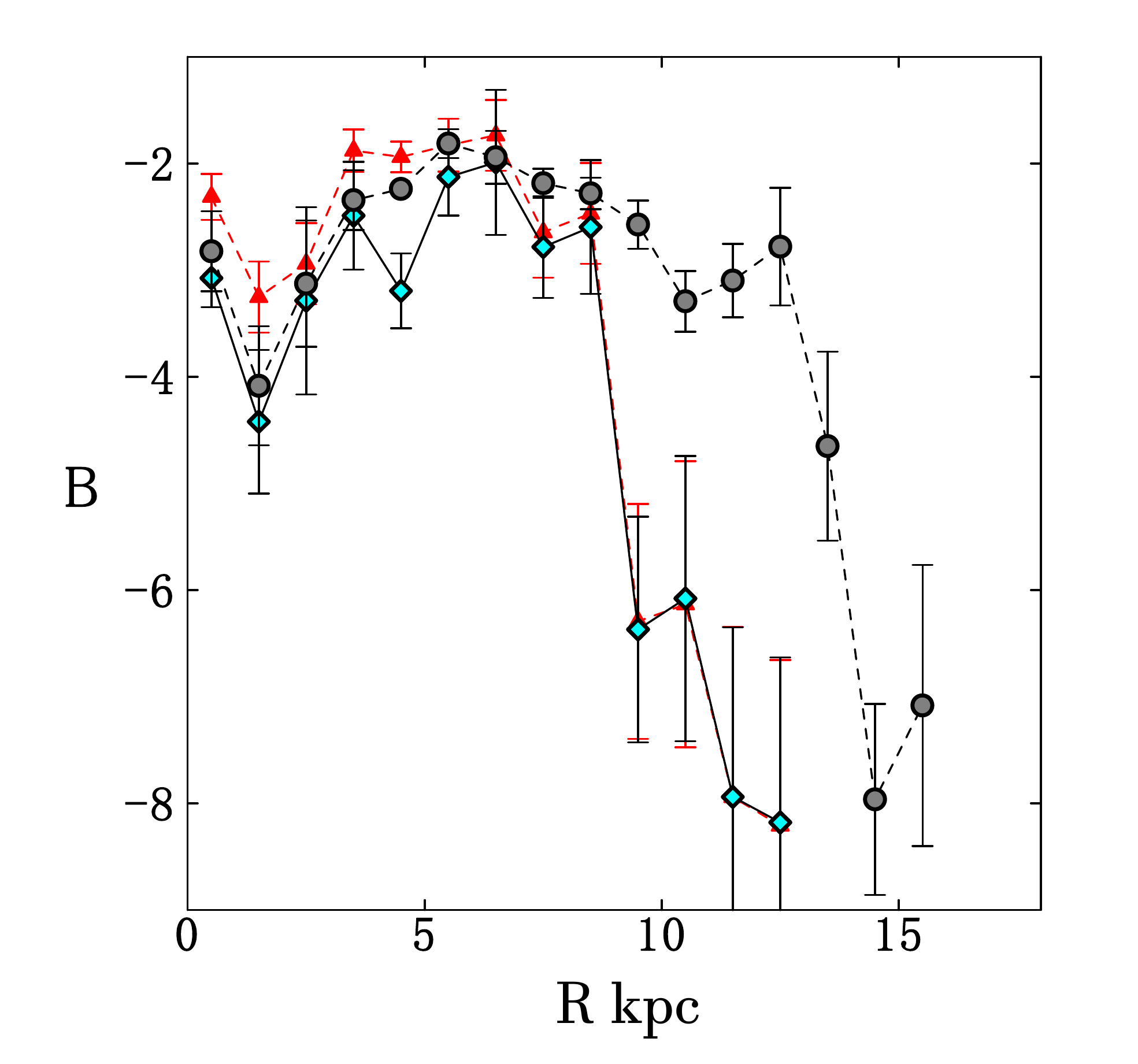}  
\hskip-3mm \includegraphics[width=6cm]{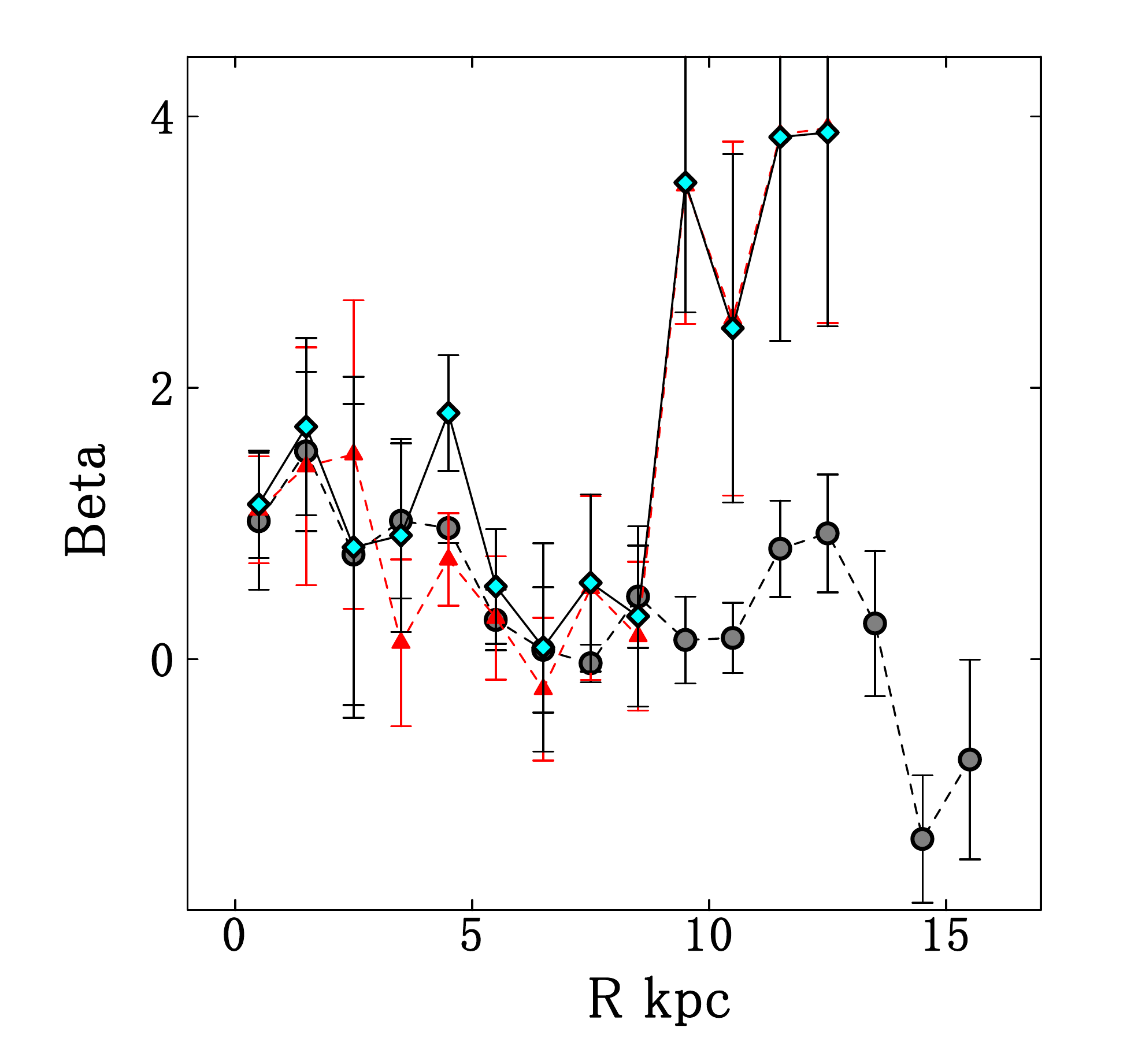}  
\hskip-3mm \includegraphics[width=6cm]{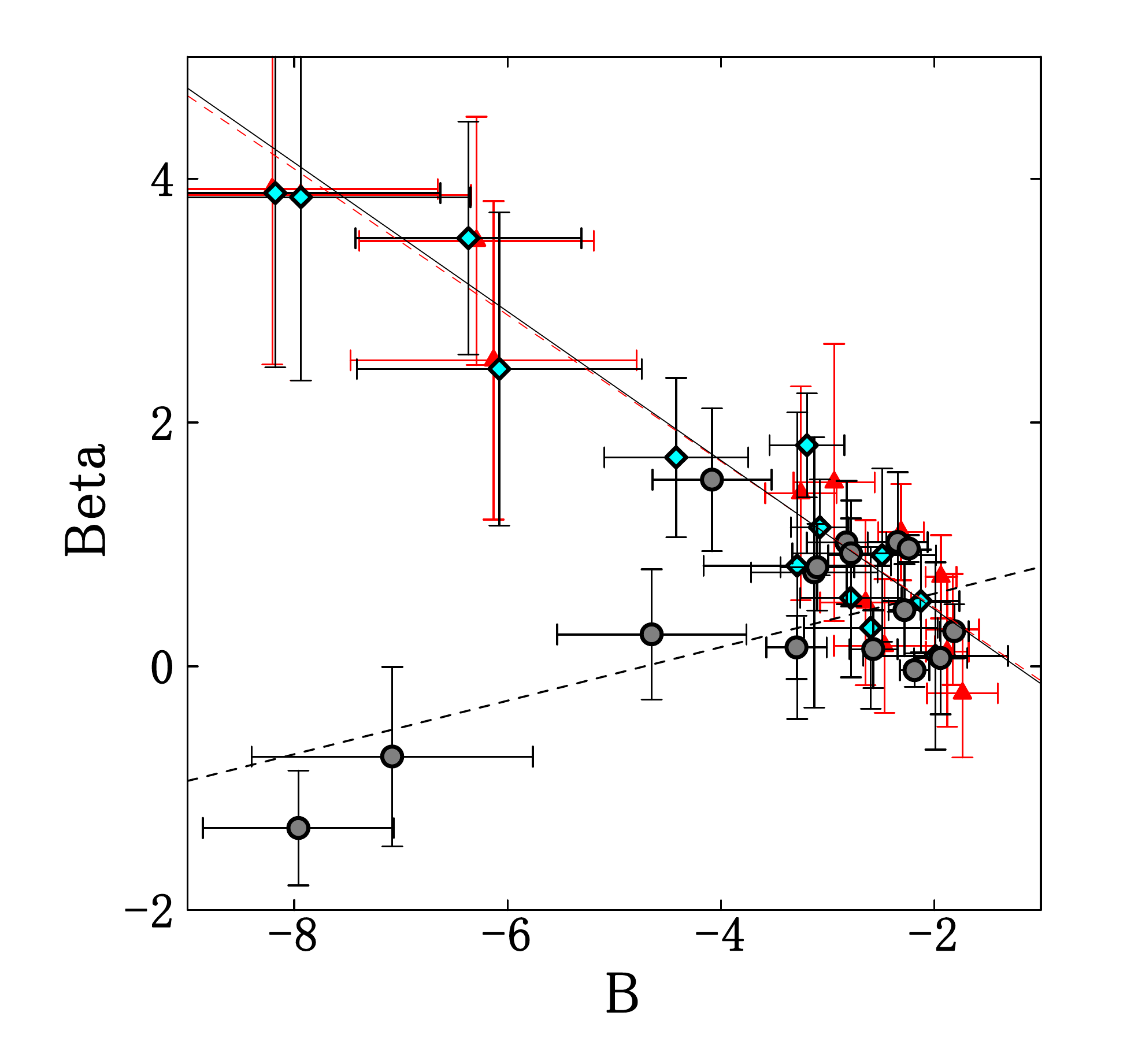} 
\end{center}
\caption{Same as figure \ref{R-A-alpha}, but for surface relations. } 
\label{R-B-beta} 

\begin{center}       
\includegraphics[width=6cm]{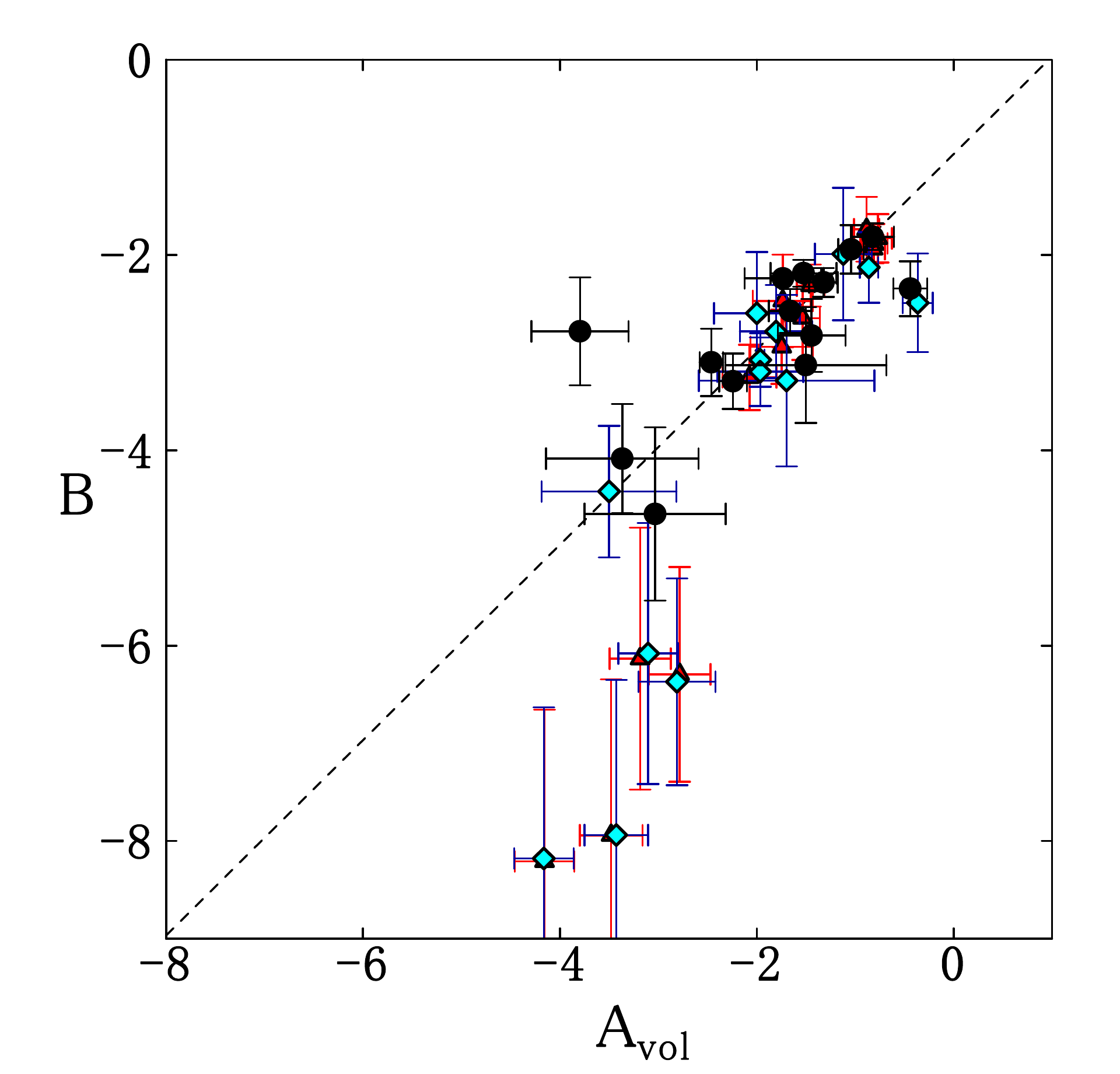}  
\includegraphics[width=6cm]{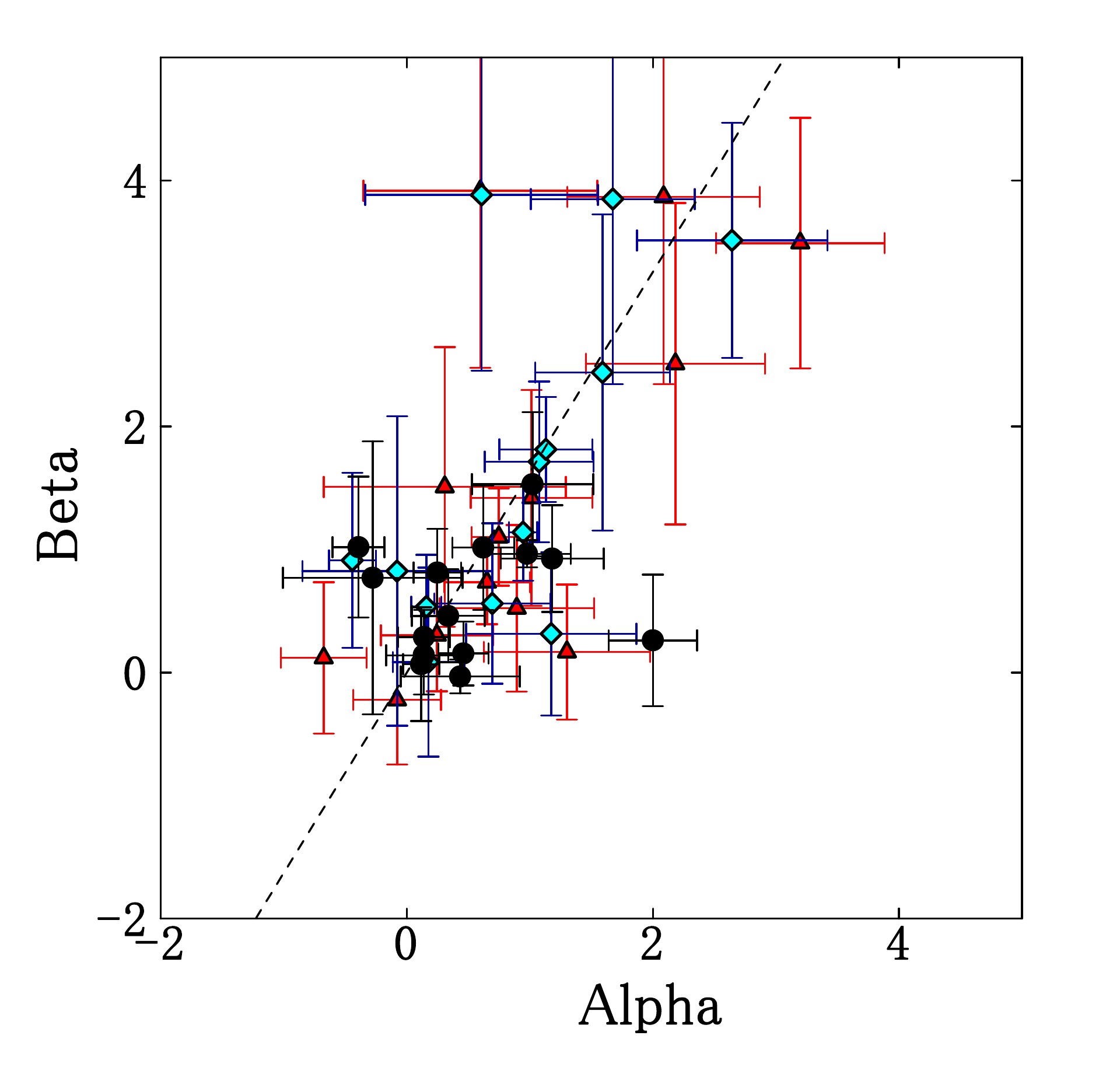}   
\end{center}
\caption{Correlation between fitted SF coefficients of surface- and volume-SF laws used in figures \ref{R-A-alpha} and \ref{R-B-beta} (left), and same between power-law indices (right). Circles, triangles, and diamonds denote \Htwo\, HI and total gases, respectively.}
\label{VScomp}  
\end{figure*}

In figures \ref{R-A-alpha} and \ref{R-B-beta} we show the SF coefficients and power-law indices as a function of the radius $R$, as obtained by the least squares fitting to the SF law plots at 1 kpc radius interval. We also show the relation between the index and coefficient.

The $\A-R$ and $B-R$ relations demonstrate and confirm that the SFR varies with the radius, revealing higher SFR in the inner region and the steep decrease toward the outer Galaxy. This decrease obviously follows the decrease of the gas density with the radius.
 
The $\alpha-R$ and $\beta-R$ relations show that the SF power-law index for molecular gas remains nearly constant within $R\sim <8$ kpc. The index for \Htwo\ gas in the volume SF law, then, increases toward the outer region. In contrast, the \Htwo\ index for the surface law seems to remain constant or even decreases. On the other hand, indices for volume and surface HI and total gases show similar behaviors, increasing toward the observed outer edge. 
 
An interesting finding in these plots is that the SF parameters, $\A$, $\alpha$, $B$, and $\beta$, are not constant at all, showing that the SF law in the Galaxy is not universal at all. The values of $\A$ and $B$ vary over three orders of magnitudes from inner to outer Galaxy. Such variable SF law may contradict the currently established universal SF law in galaxies, which requires a constant value of the coefficients $\A$ and $B$ for a given value of gas density. 
 
In figure \ref{VScomp} we plot the fitted coefficients and indices for the volume- and surface-SF laws used in figures \ref{R-A-alpha} and \ref{R-B-beta}. The plotted values are not perfectly independent because the used $\vsfr$ and $\ssfr$ are linearly related. Therefore, these plots include the correlation between the used volume- and surface-gas densities. It is readily shown that the SF coefficients $\A$ and $B$ are linearly correlated at high SF regions showing $\A=B+0.967$ by the definition of volume and surface SFR, except for low SF regions corresponding to low-density outer Galaxy. The power-law indices are weakly correlated to each other as $\beta\sim 1.63\alpha$, and are largely scattered around this relation. 

Besides the radial variation, there exists a mutual correlation between the power-law index and SF coefficient, as shown in figures \ref{R-A-alpha} and \ref{R-B-beta}. The plots reveal a clear anti-correlation between the index and coefficient. This implies that the SF law is controlled also by SFR in the sense that the higher is SFR, the flatter becomes the index. 

If the correlation can be represented by a relation as 
\begin{equation}
\alpha= \alpha_0 + c  \A,
\end{equation}
and
\begin{equation}
\beta=\beta_0+d B,
\end{equation} 
we obtain the values as listed in table \ref{tabEmp}.
Inserting these parameters in equations (\ref{eqschrho}) and (\ref{eqschsigma}), we have empirical SF laws with a single-parameter function. 
This formulation may become a generalization of the current SF law that had two parameters of the index ($\alpha$ or $\beta$) and SF coefficient ($\A$ or $B$).  
  
\begin{table} 
\caption{Empirical relation between the index and coefficient.}
\begin{center}
\begin{tabular}{ll}  
\hline
\hline
Index & Coefficient \\ 
\hline 
$\alpha_{\rm 0:mol} =-0.35\pm0.15$ & $c_{\rm mol}=-0.42\pm 0.08$ \\ 
$\alpha_{\rm 0:HI}= -0.29\pm0.22$ & $c_{\rm HI}= -0.62\pm 0.11$\\
$\alpha_{\rm 0:tot}=-0.40\pm0.15$ & $c_{\rm tot}= -0.62\pm 0.07$ \\
\hline 
$\beta_{\rm 0:mol}= 1.04\pm 0.18$ & $d_{\rm mol}=+0.22\pm 0.07$\\ 
$\beta_{\rm 0:HI}= -0.72\pm 0.36$ & $d_{\rm HI}=-0.60\pm 0.13$\\
$\beta_{\rm 0:tot}= -0.75\pm 0.47$ & $d_{\rm tot}= -0.61\pm 0.14$\\
\hline
\end{tabular}
\end{center}
\label{tabEmp}
\end{table}

\subsection{Self-regulation of SF}

The highest $\A$ and $B$, and hence the most active SF activity, is found in a radial zone at $R\sim 3-6$ kpc centered by the 4-kpc molecular ring (figures \ref{R-A-alpha}, \ref{R-B-beta}). The corresponding Schmidt plots in the panels at $R=2 \sim 6$ kpc of figures \ref{rho2kpc} and \ref{sig2kpc} exhibit flat slopes with indices less than 1.
 
Such a flat SF law may imply that the star formation in high-SFR regions tends to be saturated by molecular gas dissociation due to UV emission as well as by injection of kinetic and thermal energies by stellar winds and supernova explosions in active SF regions (Krumholz et al. 2009).  

\section{Summary and Discussion}
 
The Schmidt-Kennicutt law between the volume SFR $\vsfr$ and volume gas densities $\rho_{\rm gas}$ of HI, \Htwo, and total gases was shown to vary with the galacto-centric distance $R$. It was shown that the SF power-law index $\alpha$ is significantly smaller in the inner disc than in the outer Galaxy. Even an inverse index was found at radii near the 4-kpc molecular ring. The SF index was also found to be inversely correlated with the coefficient $\A$ of SFR, suggesting regulation of SFR by increasing SF activity. Similar results were obtained also for the surface SF laws. We, thus, suggest that the resolved SF law in the Galaxy may be neither uniform, nor does give the universal scaling relation. 
 
Variation of the SF index depending on the gaseous phase may be explained by the molecular gas regulated star formation (e.g., Krumhotz et al. 2009). If we assume that molecular clouds are formed from HI gas by a similar power law as $\rhoh2 \propto \rhohi^n$, we may expect that $\alpha({\rm HI})\sim \alpha({\rm H_2})+n>\alpha({\rm H_2})$, and so for $\beta$.
The variable SF index may be, thus, a measure of "directness" of star formation in the sense that the more additional process is involved prior to SF, the steeper becomes the index. This may also explain the steeper index of SF laws for surface gas densities than for volume densities. In other words, the $\alpha$ index is a more direct measure of star formation than $\beta$, trivially indicating that the star formation is more directly related to the volume density than to the column density.

\begin{figure} 
\begin{center}       
\includegraphics[width=7cm]{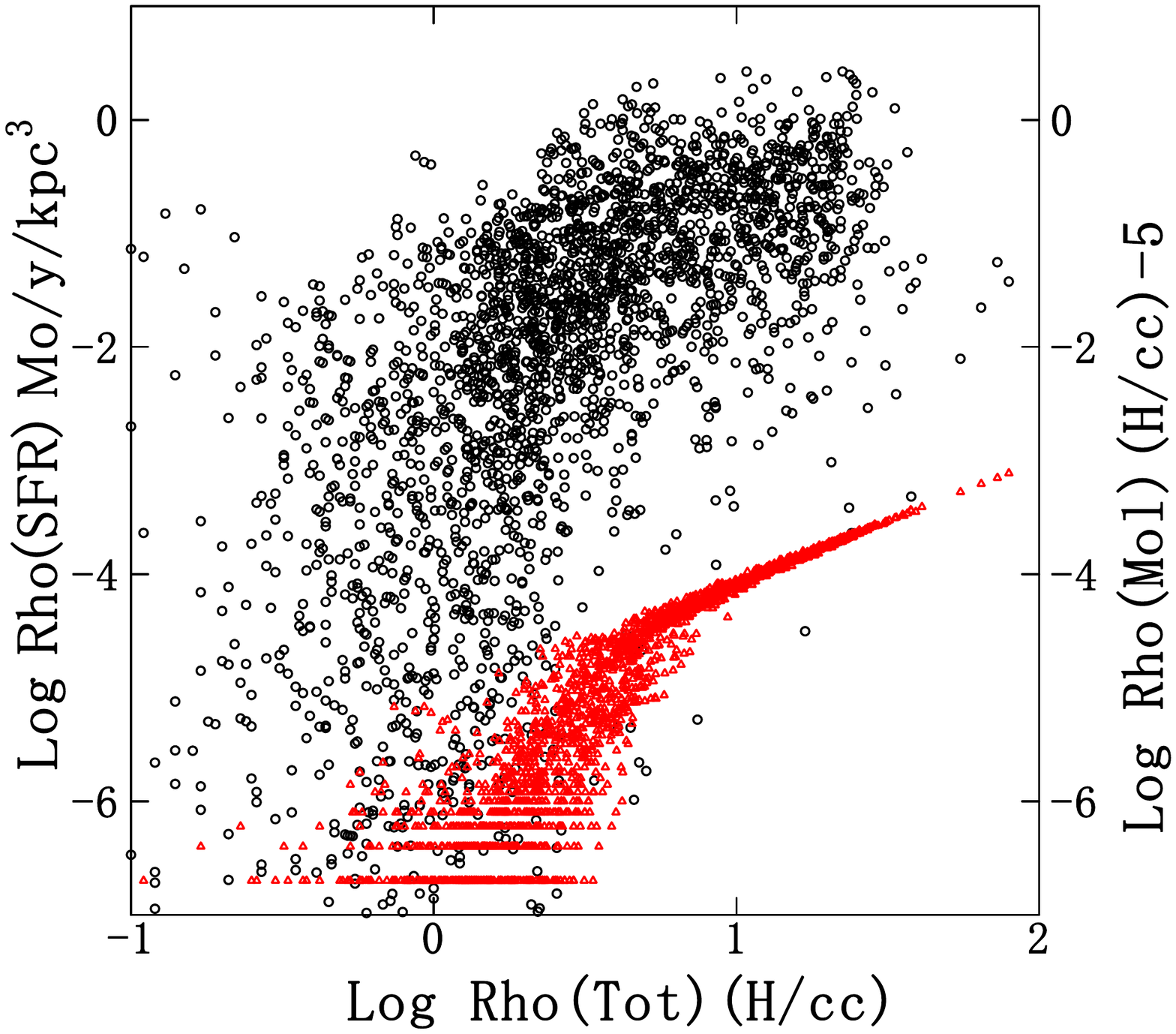}    
\end{center}
\caption{Schmidt plot of SFR against $\rhotot$ (raw data from Paper I: grey circles). Triangles show \Htwo\ gas density ($\rhoh2 \times 10^{-5})$ plotted against $\rhotot$, showing phase transition from HI to \Htwo\ around the threshold density of $\rhotot \sim 3$ H cm$^{-3}$. The bend from steep to mild index occurs around this threshold density.}
\label{trans}  
\end{figure}

This consideration may also apply to the SF law for total gas, for which the index is a mixture of indices for \Htwo\ and HI, and is steeper than the index for \Htwo\ gas. It was also found in Paper I that the SF power-law index for total gas has a bend from steep to milder around the threshold density of $\rhotot\sim 3$ H cm$^{-3}$, which we reproduce in figure \ref{trans}. In the figure we compare this SF law with the plot of molecular gas density against the total gas density, where also a bend exists. Both the bends appear at the same threshold density of $\rhotot \sim 3$ \Hcc. These bent behaviors are in agreement with the model of molecular-gas regulated star formation in galaxies (Booth et al. 2007; Bigiel et al. 2008), and the bent SF law readily found in galaxies (Krumholz et al. 2009).  
 
 The HI-to-\Htwo\ transition in low density regions, where star formation is settled from forming \Htwo\ gas, makes the SF index steeper than that in the molecular-gas dominant regions. The steeper index in the outer Galaxy with lower density and lower molecular fraction reflects such indirect star formation, where the index tends to become larger because of the additional processes of transition from HI to \Htwo\ and formation of molecular clouds.    
  
The $\vsfr - \rhoh2$ law is, thus, considered to be a direct measure of star formation (Krumhotz et al. 2008). The other relations for $\rhohi,\ \rhotot,\ \sighi$, and $\sigtot$ must be analyzed more carefully by taking into account the intervening processes such as the phase-transition between the \Htwo\ and HI, and the formation and dissociation of molecular clouds.  
The current studies used to use the surface relations, which implicitly contain the indirect or unrelated HI gas by projecting on the galactic plane.   \\

\noindent{\it Acknowledgements}: The author thanks Dr. H. Nakanishi of the Kagoshima University for the HI and \Htwo\ 3D cube data.

 \newpage
 \appendix
 \section{Volume- and surface-SF laws in annulus rings.}

\begin{figure*} 
\begin{center}       
 \includegraphics[width=6cm]{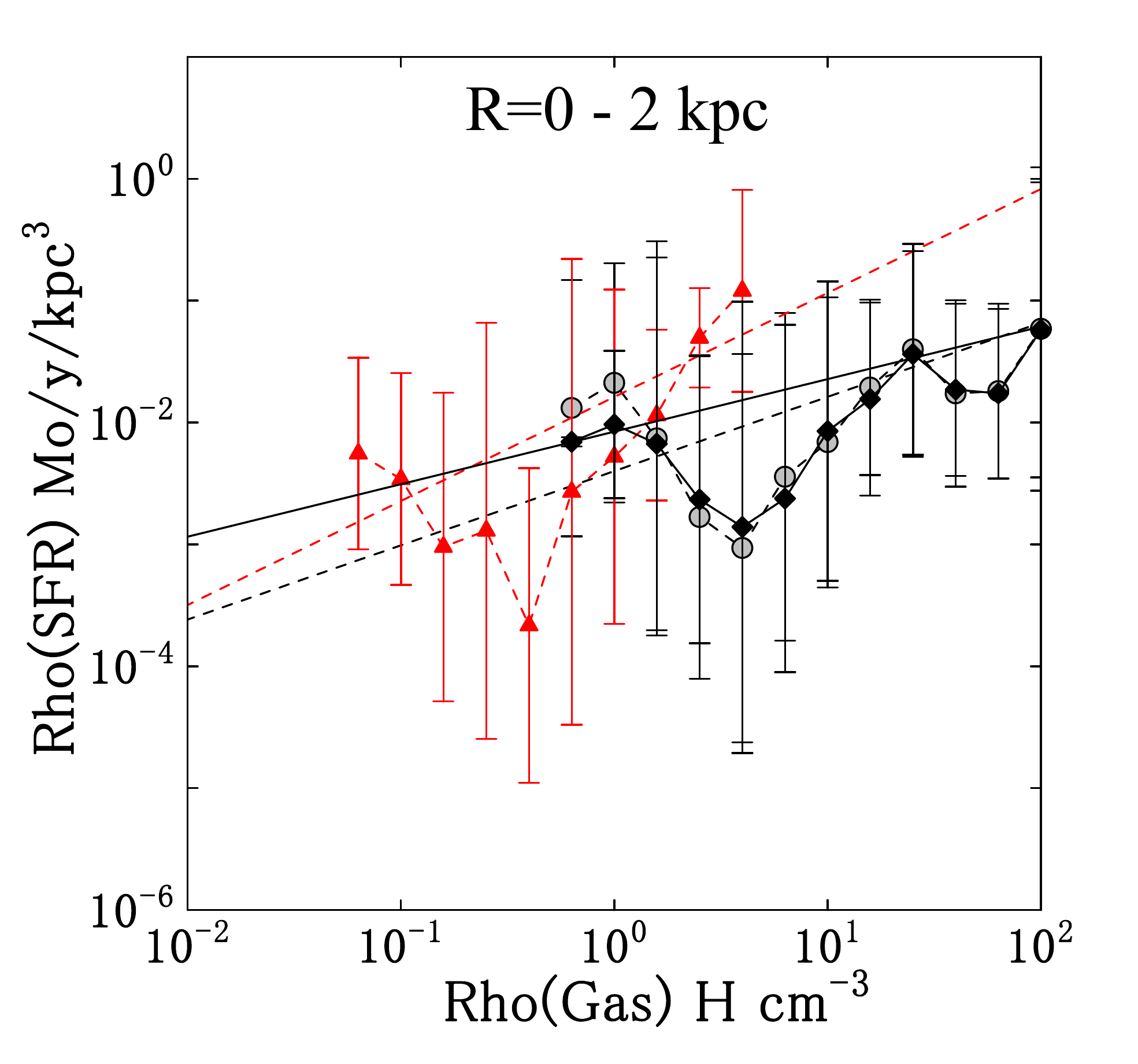}  
\includegraphics[width=6cm]{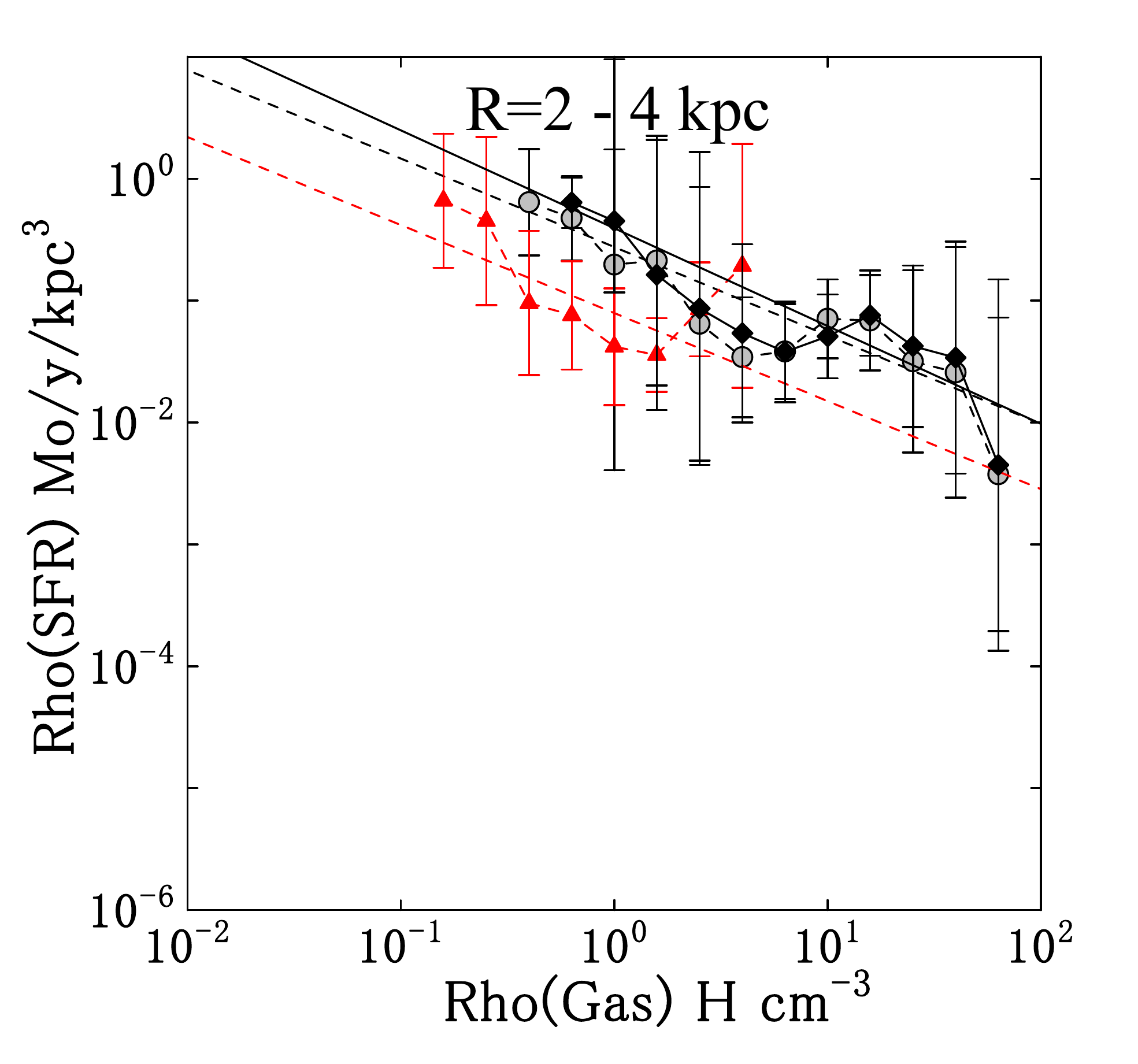}  \\
\includegraphics[width=6cm]{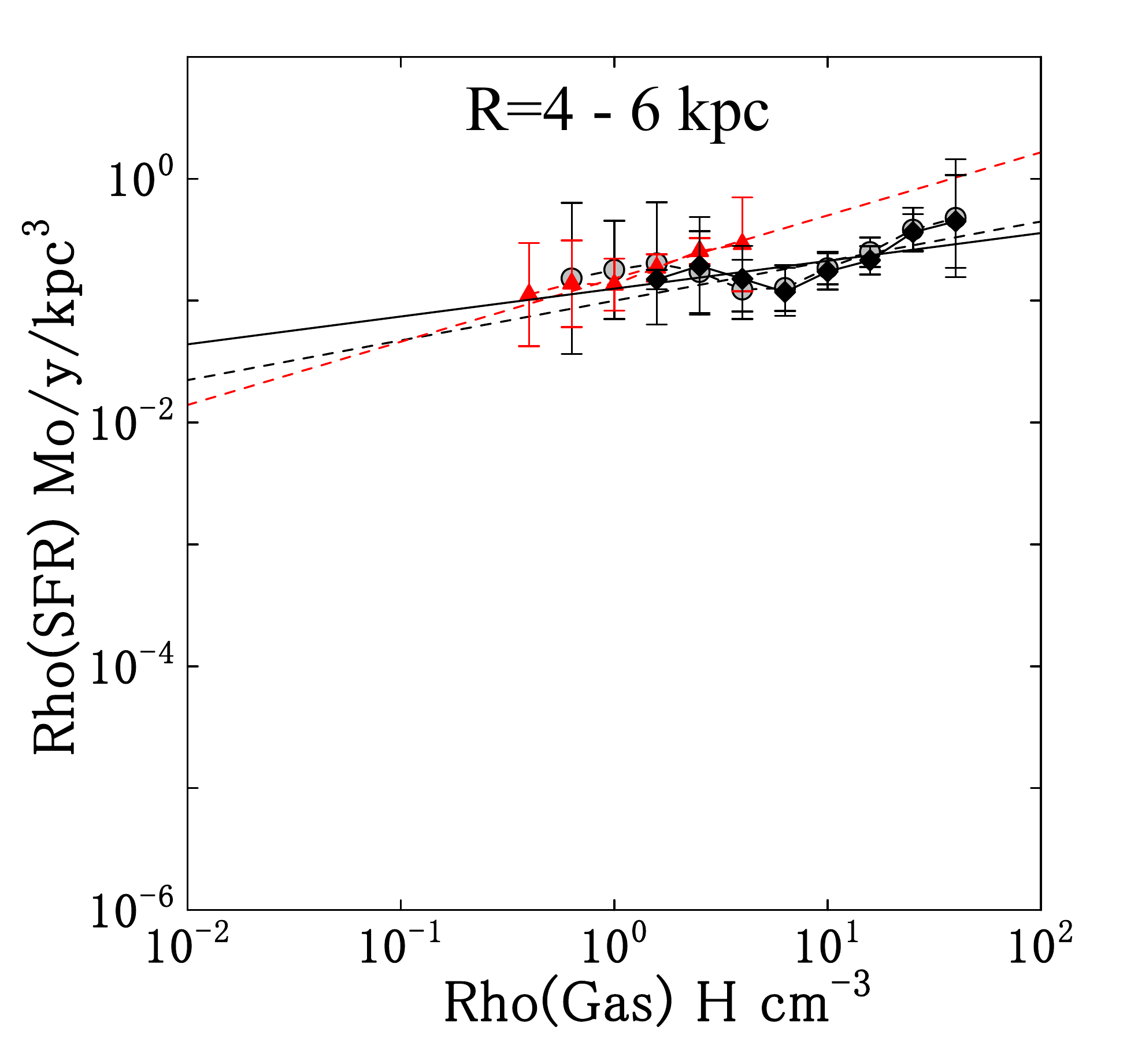}  
\includegraphics[width=6cm]{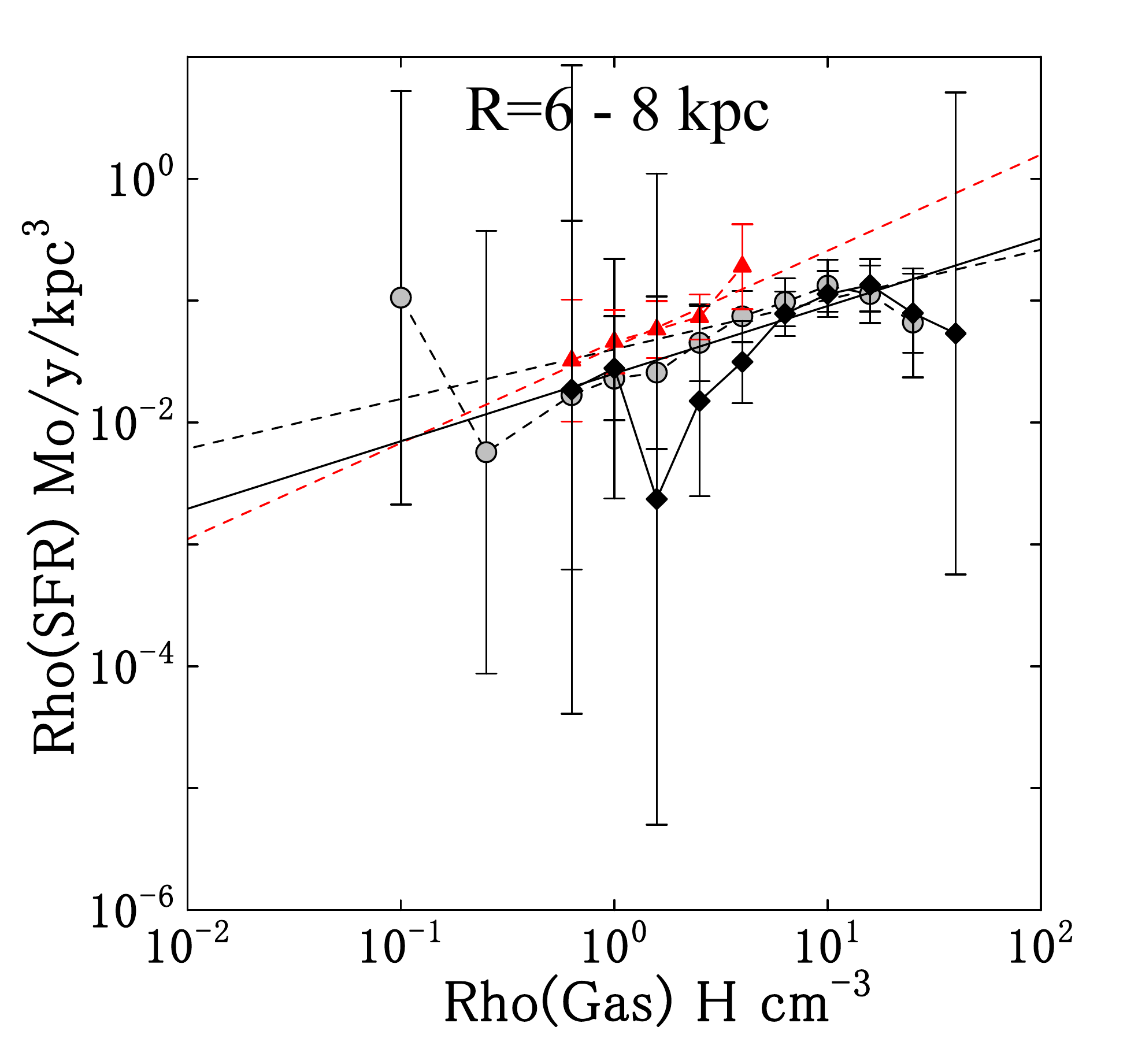}  \\
\includegraphics[width=6cm]{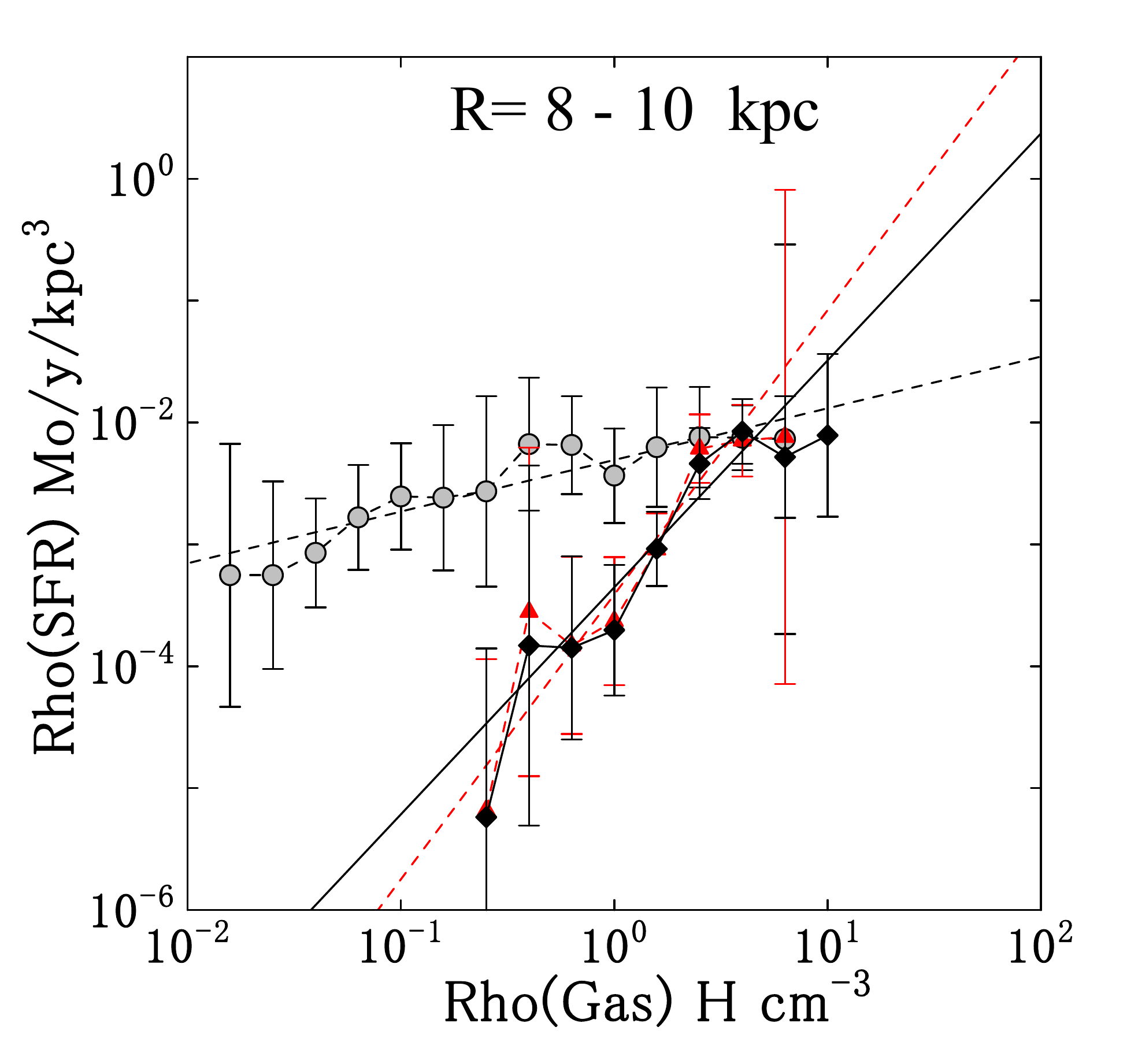}  
\includegraphics[width=6cm]{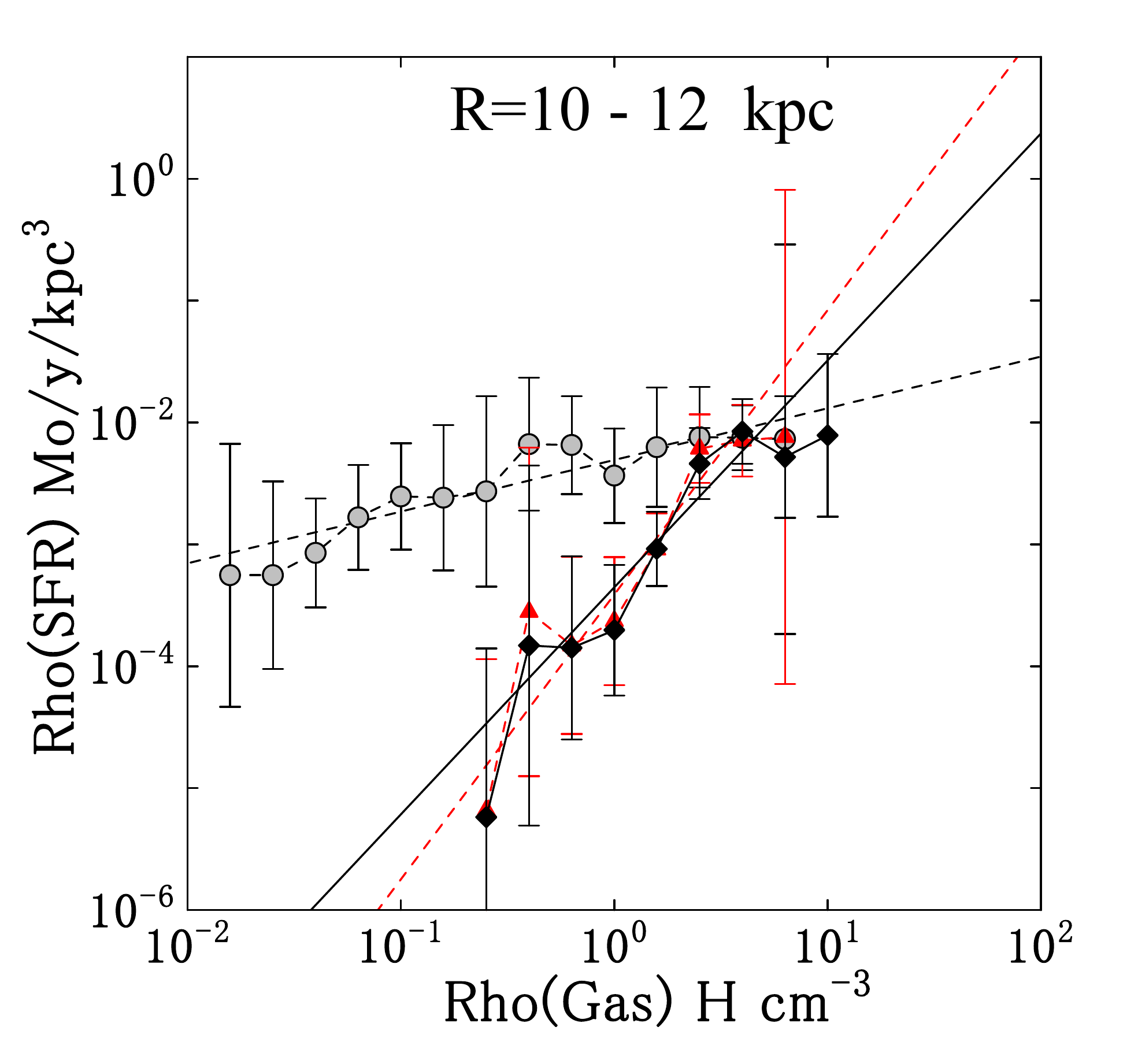}  \\
\includegraphics[width=6cm]{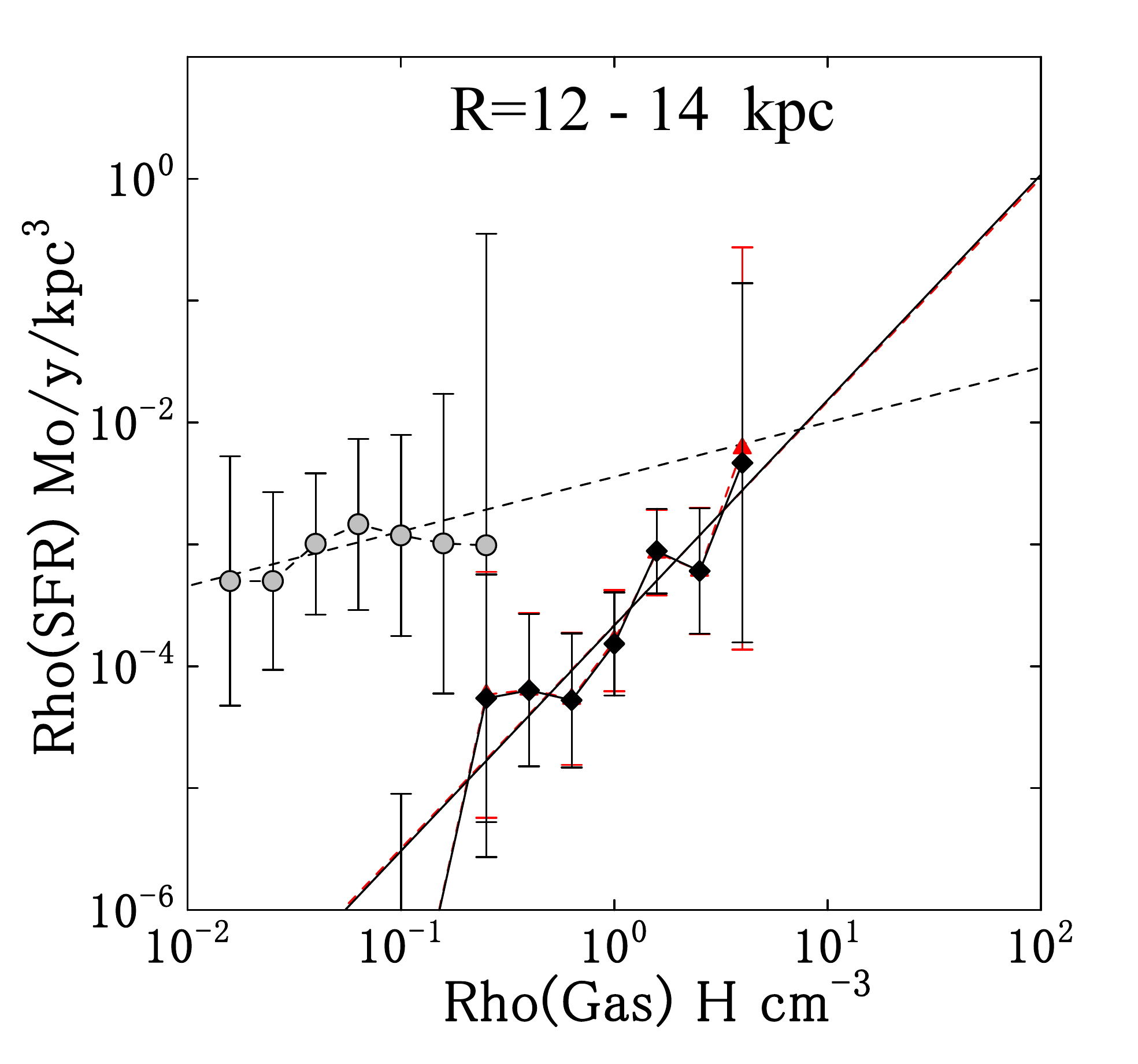}  
\includegraphics[width=6cm]{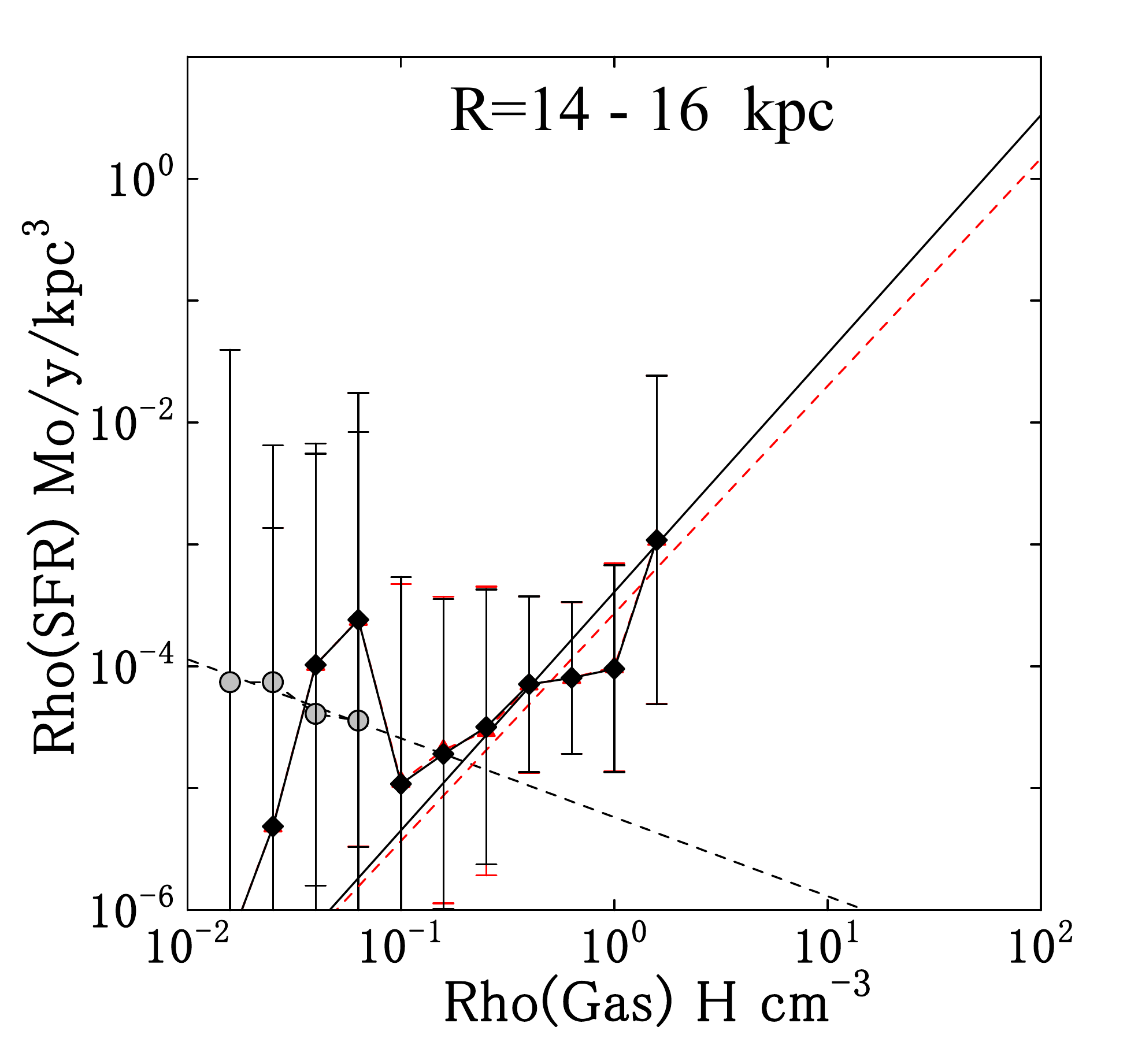}  
\end{center}
\caption{Radial variation of volume SF law at different radii plotted every 2 kpc interval. Circles, triangles, and diamonds denote $\vsfr$ for volume densities of \Htwo\, HI and total gases, respectively.}
\label{rho2kpc}   
\end{figure*}

\begin{figure*}  
\begin{center}       
\includegraphics[width=6cm]{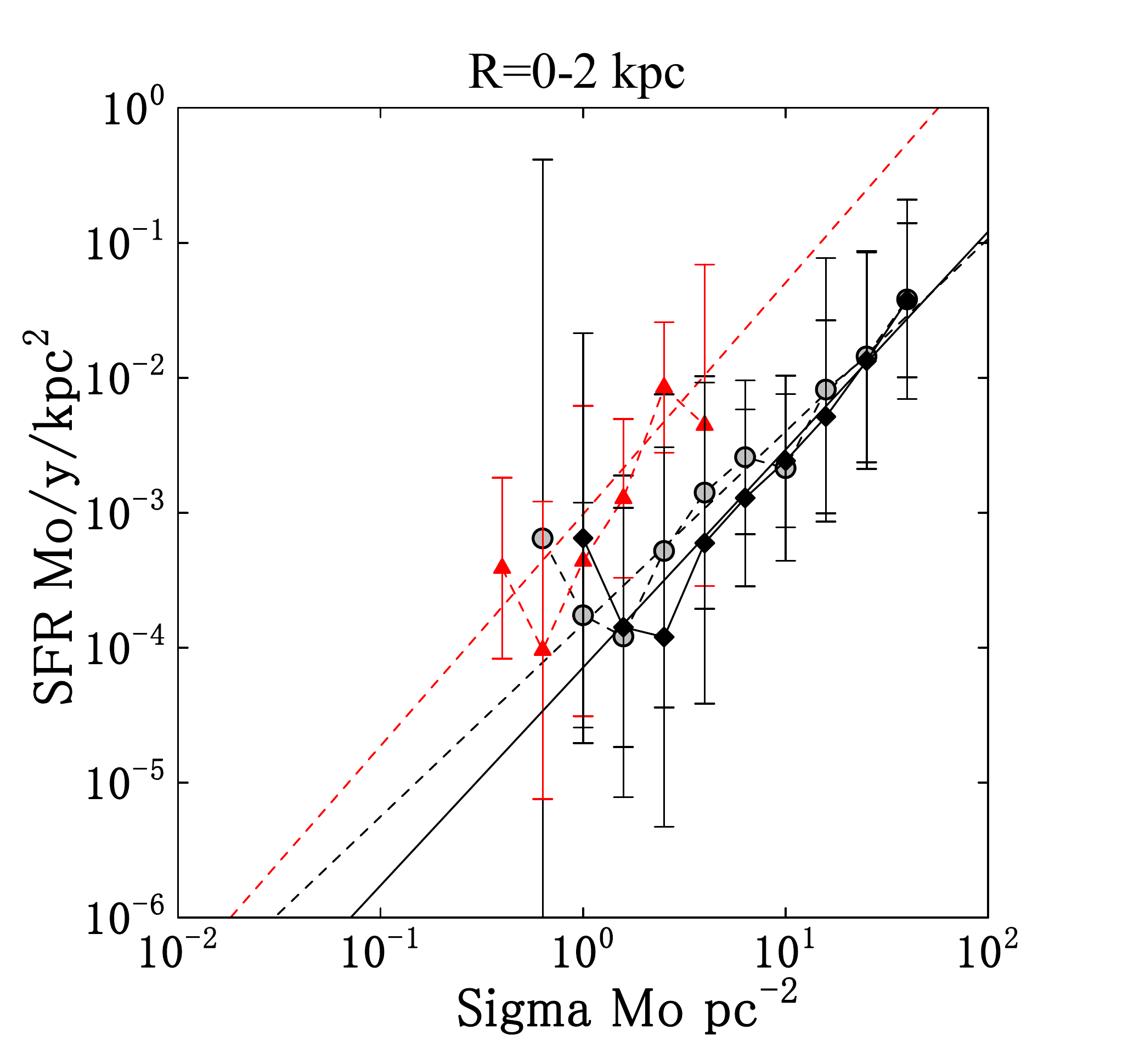}  
\includegraphics[width=6cm]{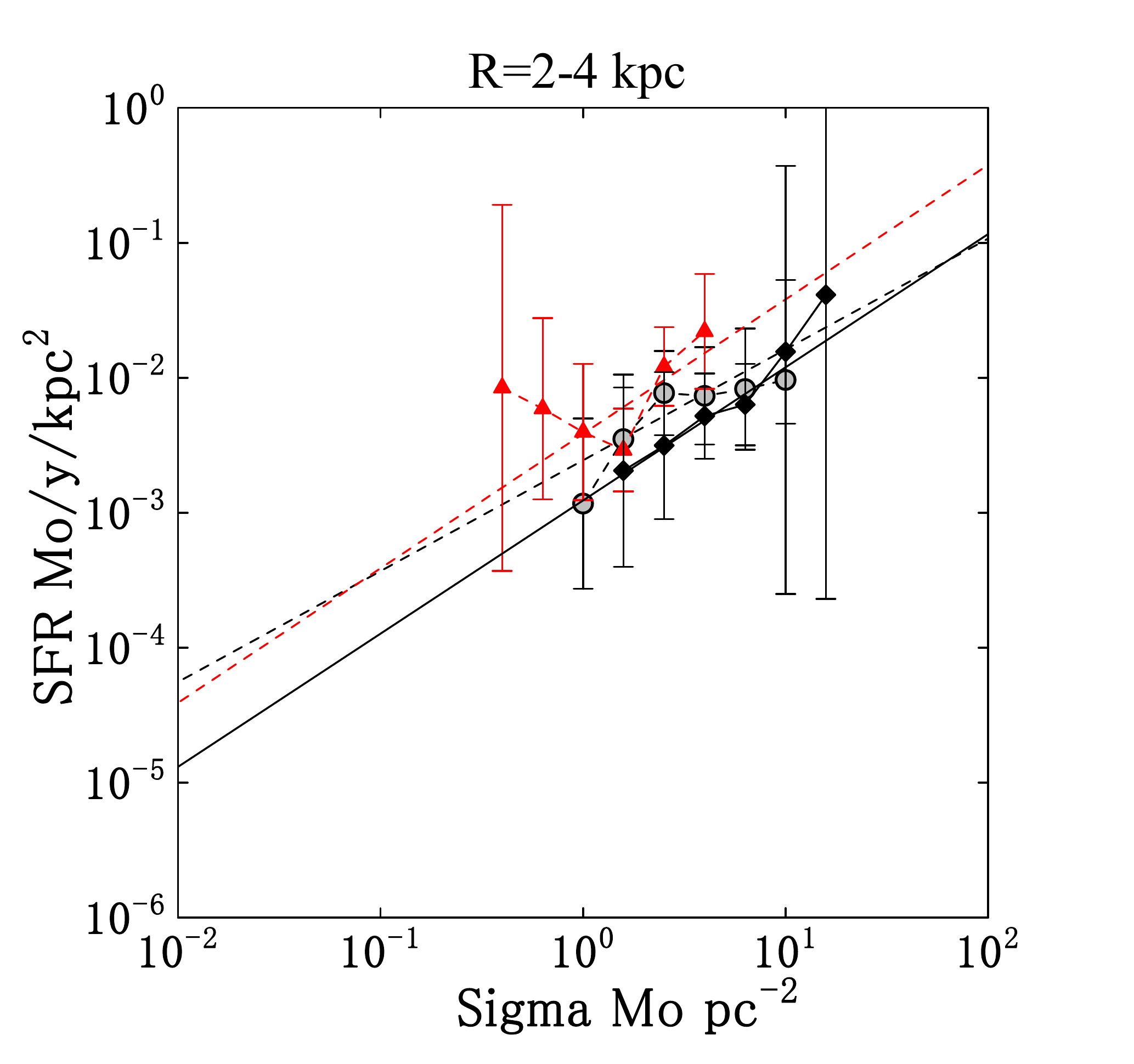}  \\
\includegraphics[width=6cm]{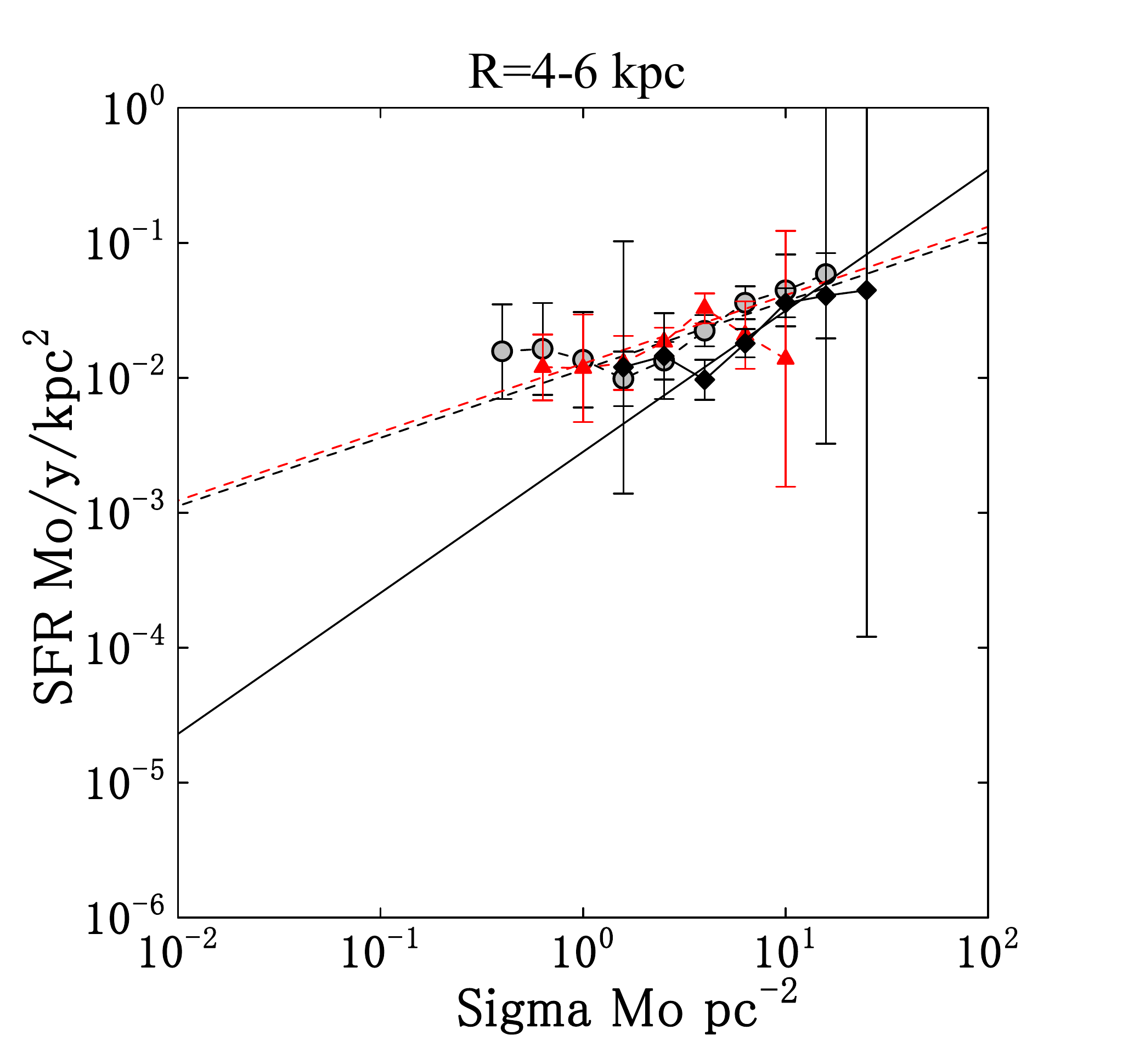}  
\includegraphics[width=6cm]{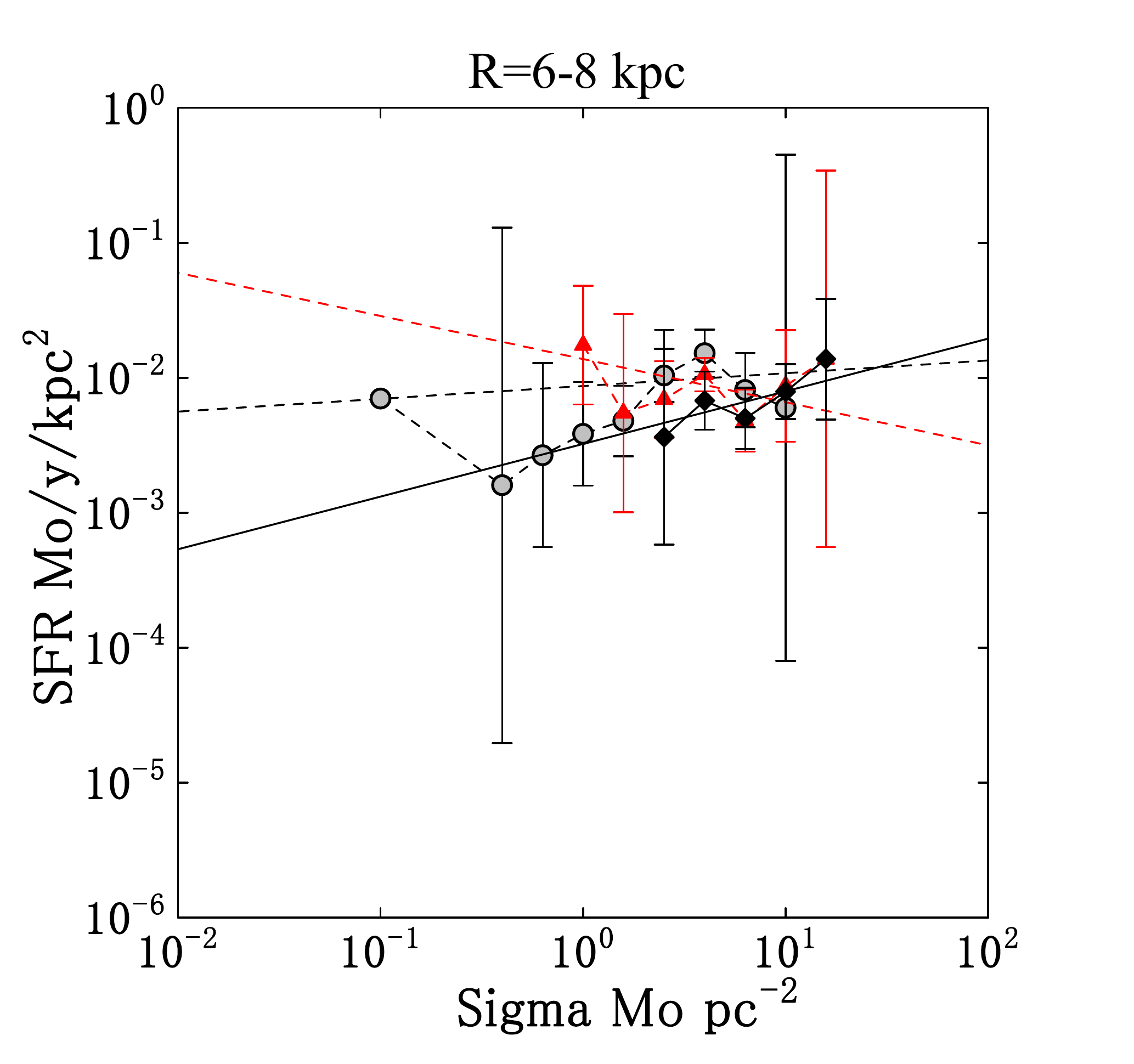}  \\
\includegraphics[width=6cm]{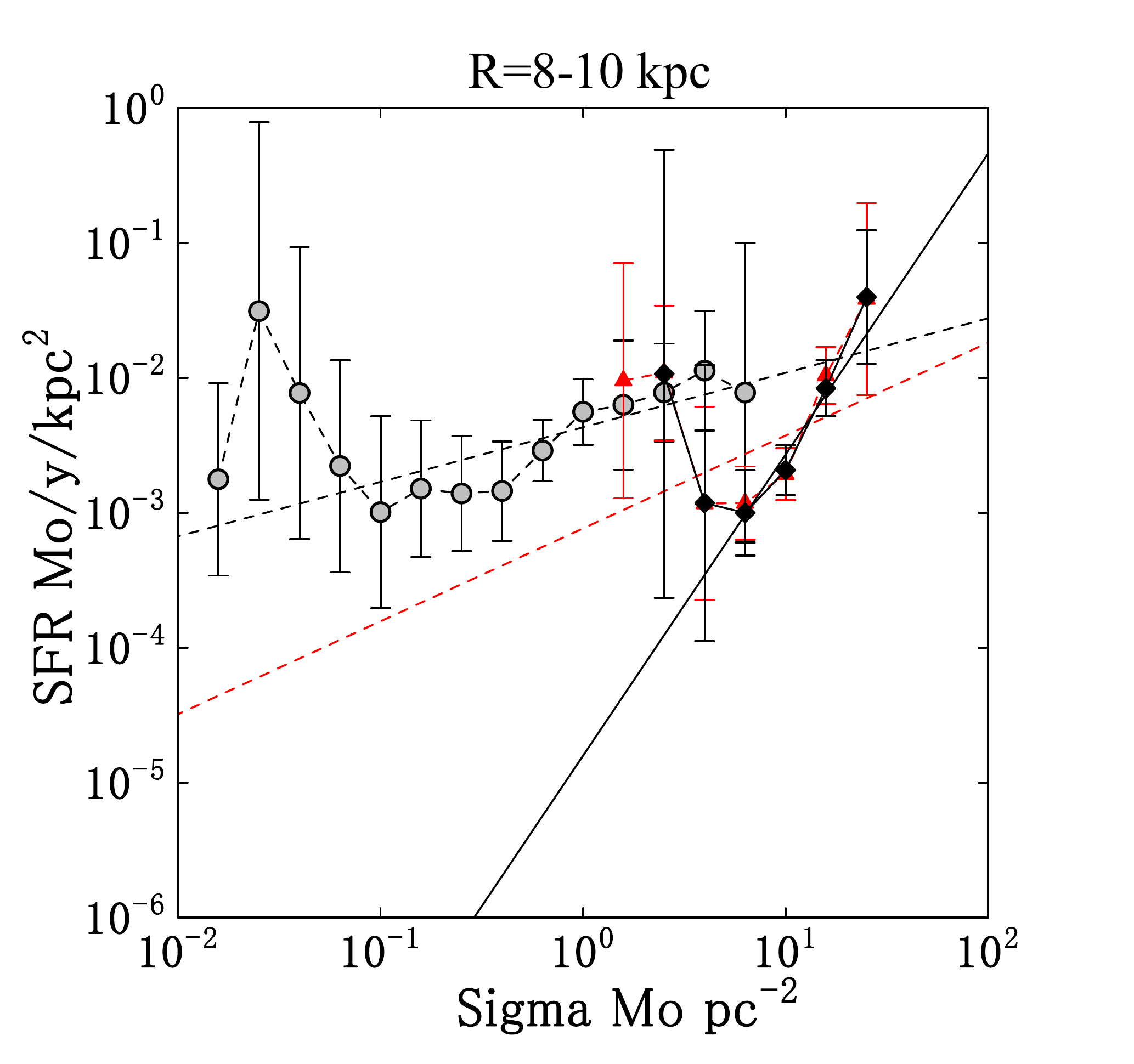}  
\includegraphics[width=6cm]{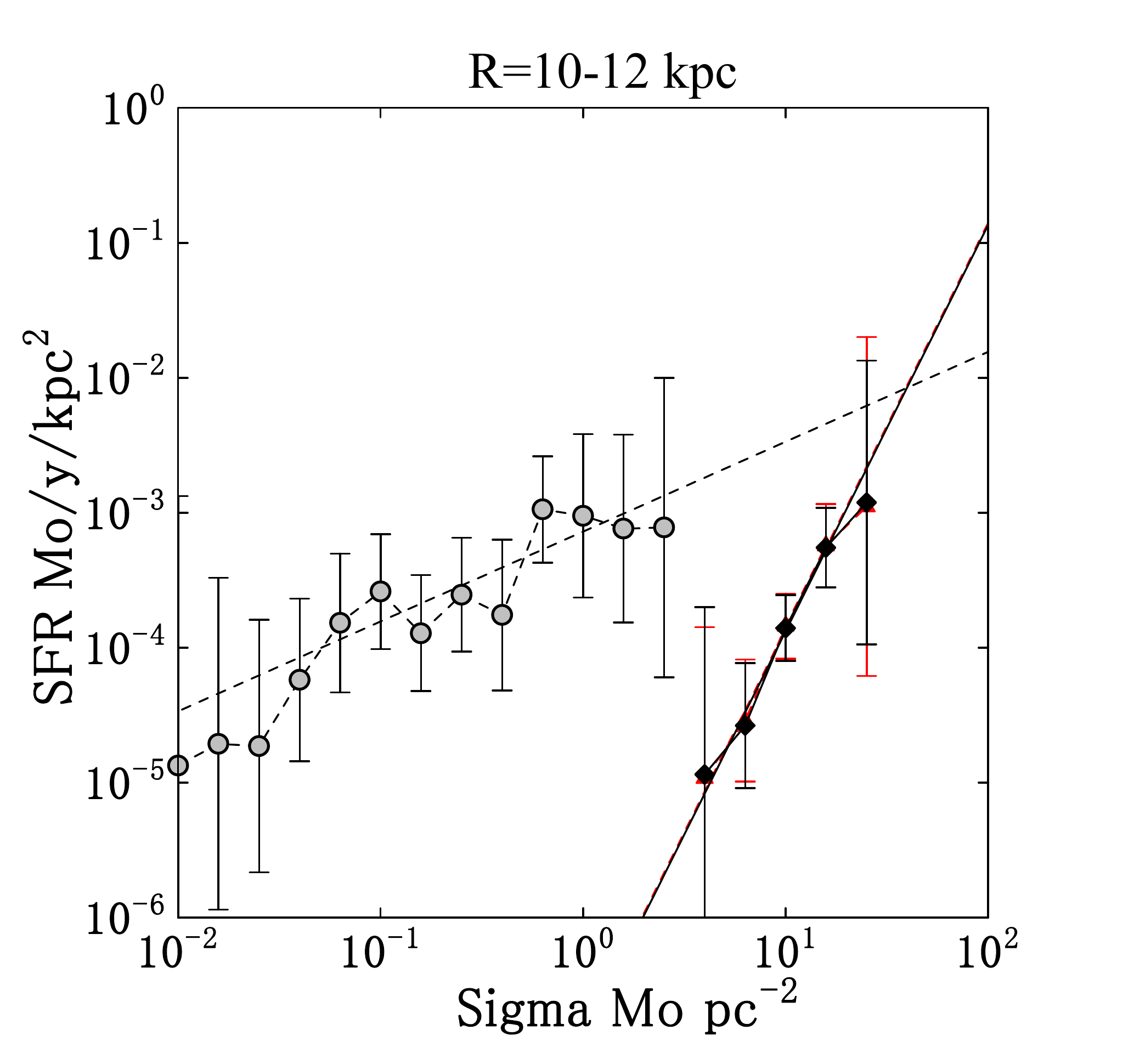}\\  
\includegraphics[width=6cm]{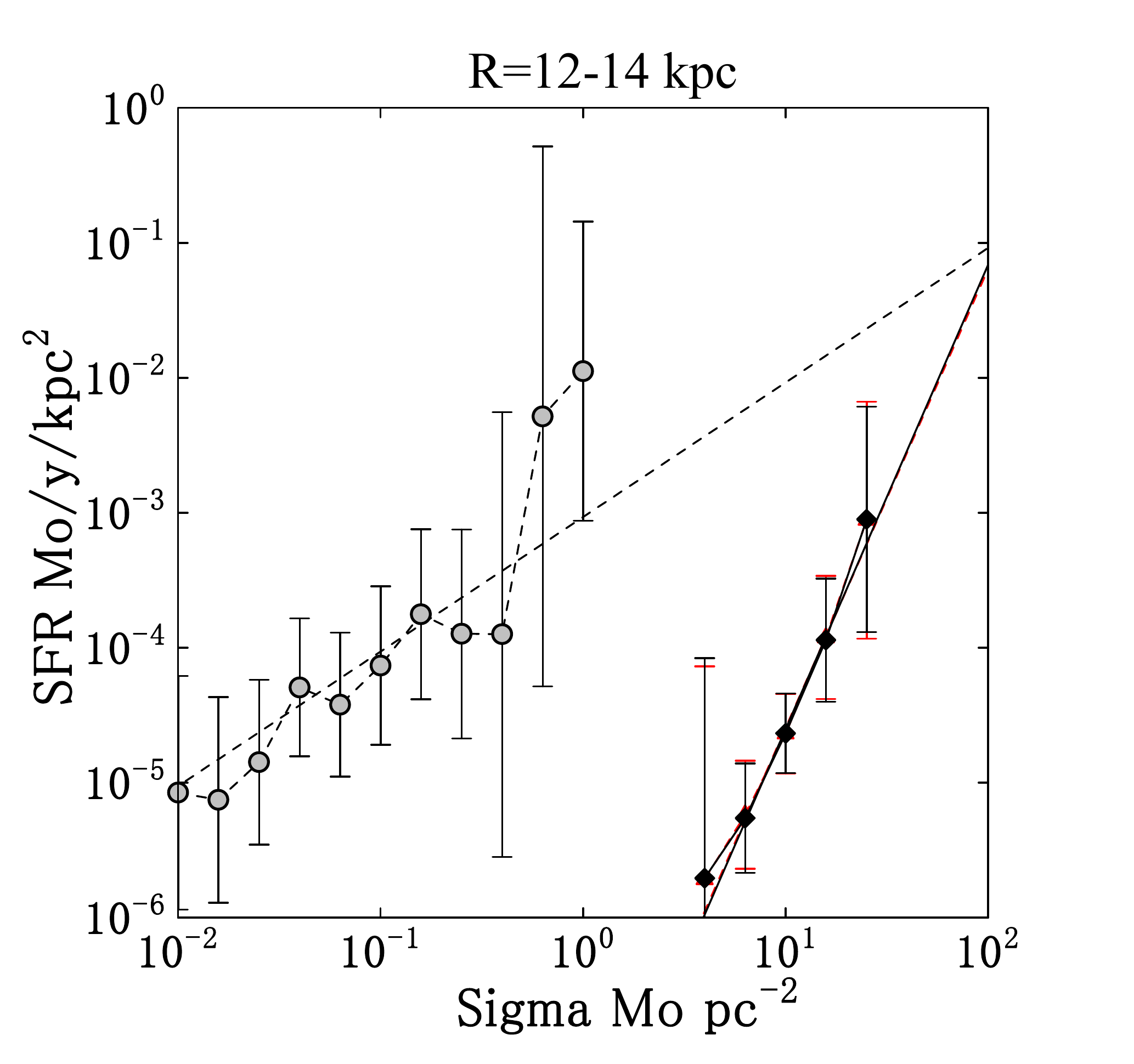}  
\includegraphics[width=6cm]{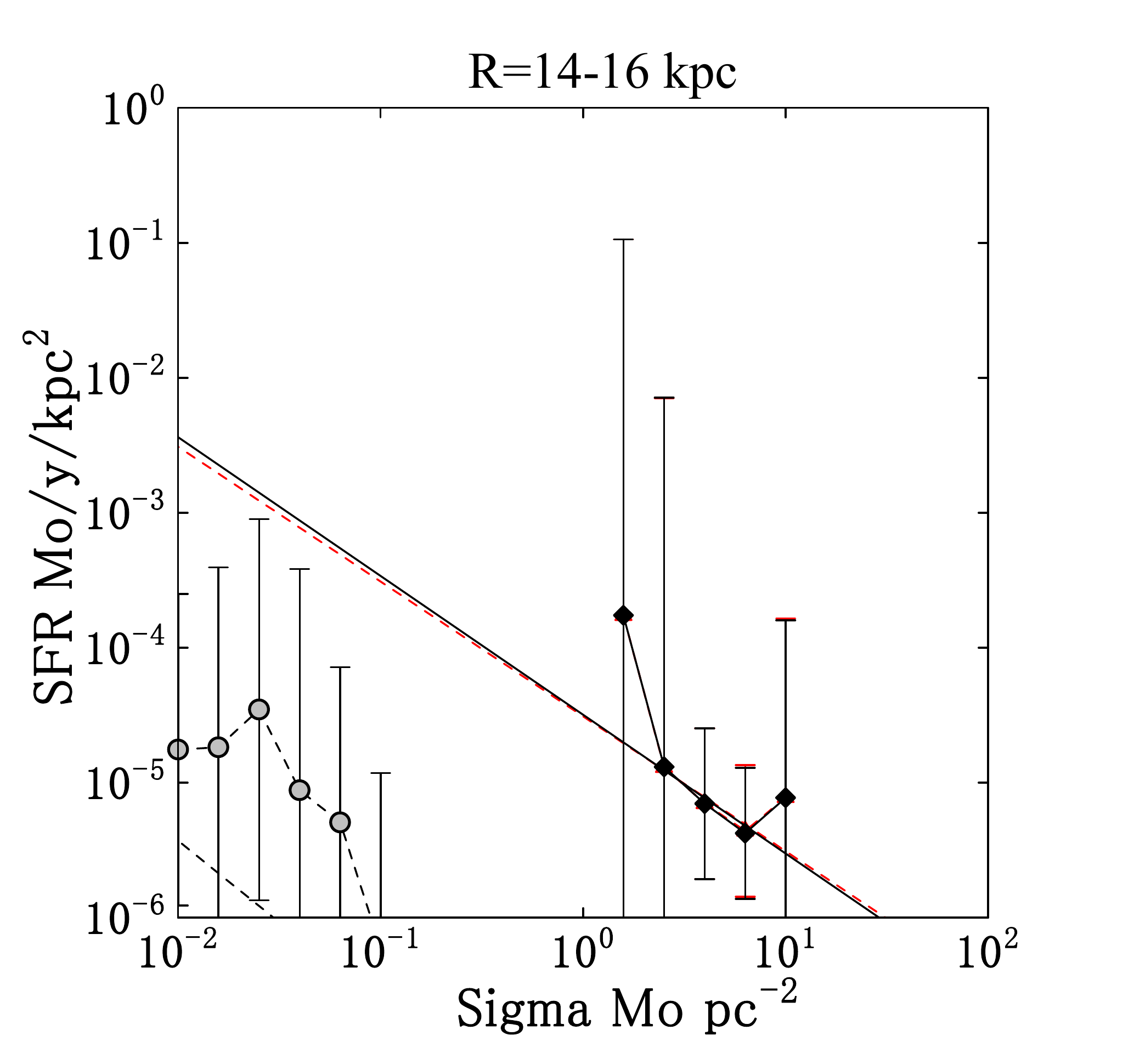}  
\end{center}
\caption{Same as figure \ref{rho2kpc}, but for the surface SFR.  Circles, triangles, and diamonds denote $\ssfr$ for surface densities of \Htwo\, HI and total gases, respectively}
\label{sig2kpc}  
\end{figure*}

In this appendix we show volume-SF laws (plots of $\vsfr$ against $\vgas$) made in individual annulus rings of 2 kpc width in figure \ref{rho2kpc}, and surface-SF laws ($\ssfr$ against $\sgas$) in figure \ref{sig2kpc}. The inserted lines show the least-squares fitting results, and the parameters (indices and coefficients) are listed in tables \ref{tab1} and \ref{tab2}.

\end{document}